\PassOptionsToPackage{modulo}{lineno}

\documentclass[
pre,
 reprint,
 amsmath,amssymb,
 aps,
 pre,
]{revtex4-2}

\usepackage{graphicx}% Include figure files
\usepackage{dcolumn}% Align table columns on decimal point
\usepackage{bm}% bold math
\usepackage{cancel}
\usepackage{hyperref}

\usepackage[dvipsnames]{xcolor}
\definecolor{mycolor}{rgb}{1, 0, 0} % Use 1,0,0 for red and 0,0,0 for black

\usepackage[normalem]{ulem} % Strikeout cmd \sout
\newcommand{\removed}[1]{}
\newcommand{\added}[1]{#1}

\begin{document}

\title{Dynamics of Individual Active Elastic Filaments with Chiral Self-Propulsion}

\author{Chanania Steinbock}
\email{chanania.steinbock@jhu.edu}
\author{Daniel A. Beller}
\email{d.a.beller@jhu.edu}
\affiliation{Department of Physics and Astronomy, Johns Hopkins University, Baltimore, Maryland 21218,
USA}

\date{\today}

\begin{abstract}
We study the over-damped dynamics of individual one-dimensional elastic filaments subjected to a chiral active force which propels each point of the filament at a fixed angle relative to the tangent vector of the filament at that point. Such a model is a reasonable starting point for describing the behavior of polymers such as microtubules in gliding assay experiments. We derive sixth-order nonlinear coupled partial differential equations for the intrinsic properties of the filament, namely, its curvature and metric, and show that these equations are capable of supporting multiple different stationary solutions in a co-moving frame, i.e.\ that chiral active elastic filaments exhibit  dynamic multi-stability in their shapes. A linear stability analysis of these solutions is carried out to determine which solutions are stable and a brief analysis of the time-dependent approach to stationary shape is considered. Finally, simulations are presented which confirm many of our predictions while also revealing additional complexity.
\end{abstract}

\maketitle

%\tableofcontents

\section{\label{sec:Intro}Introduction}

One-dimensional elastic filaments can experience a range of deformed states including stretched, bent, and twisted \cite{LandauLifshitzElasticityBook}. Such deformations are relevant at macroscopic as well as microscopic scales, with thermal, mechanical, hydrodynamic, and electrostatic forces exciting these modes in polymers, including biopolymers \cite{williams2000electrostatic,cosentino2005hydrodynamic,kumar2010biomolecules}. Cytoskeletal filaments, chiefly actin, microtubules, and intermediate filaments, or their homologs in bacteria, are essential to processes such as cell division, cell motility, and molecular cargo transport. As life is inherently far from equilibrium, there is great interest in studying the mechanics and dynamics of elastic filaments driven by non-equilibrium forces, especially the action of motor proteins such as kinesin, dynein, and myosin \cite{blanchoin2014actin, Hawkins2010, gittes1996directional,wu2011effects,murrell2012f,shi2020chiral,ng2025active}.

% Accordingly, when driven out-of-equilibrium by external forces, the shape and characteristics of such filaments can be quite intricate. 
% Important physical examples of such filaments include polymers in general and biological protein filaments such as microtubules and actin in particular.

A well-established experimental platform for the study of cytoskeletal filament dynamics and mechanics, in the presence of molecular motors,  is the gliding assay \cite{HowardBook, Amos1991, Weiss1991, Bourdieu1995, Kawamura2008, Liu2011, Kabir2012, Sumino2012, Scharrel2014, Inoue2015, Saito2017, Kim2018, Farhadi2018, Tanida2020, Afroze2021, Memarian2021, Zhou2022}. In these experiments, biopolymers are polymerized in vitro and sedimented onto a substrate coated in motor proteins, which propel the biopolymers over the surface. The wide range of filament morphologies exhibited in gliding assays is especially interesting in the case of microtubules, whose millimeter-scale persistence length makes large deformations unlikely to result from  thermal fluctuations alone \cite{pampaloni2006thermal}.  Microtubules in gliding assays exhibit a variety of  behaviors including propagation along straight and curved paths \cite{Amos1991, Weiss1991, Kawamura2008, Scharrel2014}, rotating around pinned heads \cite{Amos1991, Weiss1991, Bourdieu1995, Liu2011}, getting trapped in small circular loops \cite{Amos1991, Weiss1991, Kawamura2008, Liu2011, Kabir2012} and spiraling \cite{Weiss1991, Bourdieu1995, Kawamura2008, Kabir2012}. In addition, at sufficiently high surface densities, they have a wealth of  active matter behaviors, such as states with rotating nematic or polar order \cite{Inoue2015, Saito2017, Kim2018, Farhadi2018, Tanida2020, Memarian2021, Zhou2022}, lane formation \cite{Sumino2012, Inoue2015, Farhadi2018, Memarian2021}, vortices \cite{Sumino2012, Afroze2021} and clusters \cite{Tanida2020, Afroze2021}. Untangling which of these  phenomena are emergent due to collective interactions and which can be explained in terms of the behavior of individual filaments is a highly non-trivial question. 

Microtubules are hollow, cylindrical macromolecules, typically microns in length and 30 nanometers in diameter,  composed of dimers of tubulin proteins organized in parallel protofilaments which run along the length of the tube, with a well-defined polarity \cite{HowardBook}. In the cellular cytoskeleton, motor proteins such as kinesin and dynein transport cargo by ``walking'' along the microtubules, typically following an individual protofilament. The direction of travel is consistent, with kinesin-1,2,3,4,5 and 8 traveling from the negative end to the positive end while kinesin-14 and dynein travel the reverse.  Protofilaments may spiral around the microtubule in parallel helices, a phenomenon termed ``supertwist''; typical in vitro preparations contain distributions of microtubules with 12, 13, or 14 protofilaments, which respectively have right-handed, straight, or left-handed protofilament supertwist. As motor proteins travel along the protofilaments, they too spiral around the protofilament; additionally, motor proteins may occasionally side-step from one protofilament to the next \cite{Meissner2024}. The combination of these effects gives the motor proteins a primarily longitudinal velocity along the microtubule with a small secondary azimuthal velocity around the microtubule. 

\begin{figure}
    \centering
    
    \includegraphics[width=1.0\linewidth]{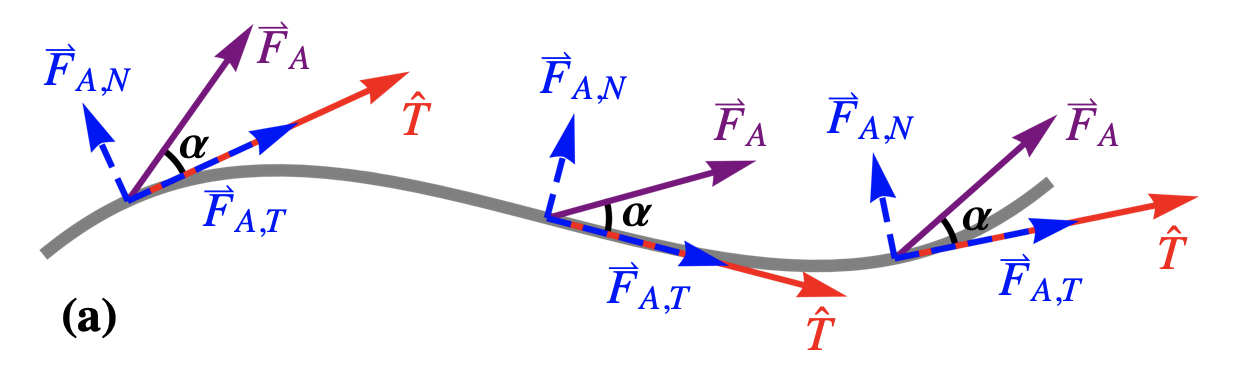}
    
    \vspace{0.2cm}
    
    \includegraphics[width=0.49\linewidth]{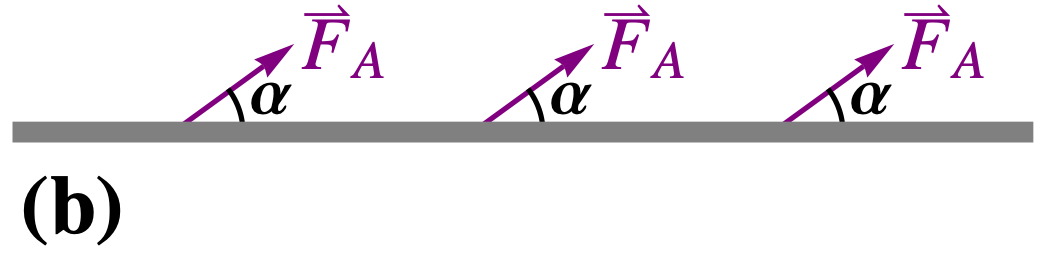}
    \includegraphics[width=0.49\linewidth]{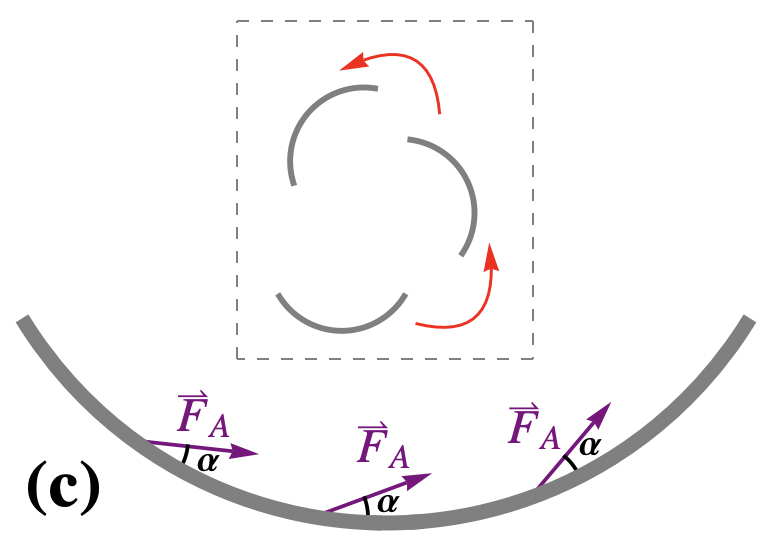}
    
    \caption{\textbf{(a)} Schematic illustration of a one-dimensional filament, such as a microtubule, experiencing a chiral active force $\vec{F}_A$. At each point along the curved filament, the active force $\vec{F}_A$ points at an angle $\alpha$ counter-clockwise from the unit tangent vector $\hat{T}$ at that point. This active force can then be decomposed into its tangential component $\vec{F}_{A,T}$ and its normal component $\vec{F}_{A,N}$.
    \textbf{(b)} Schematic representation of a microtubule in a straight conformation. Such a microtubule will simply translate across the surface. \textbf{(c)} Schematic representation of a microtubule in a curved conformation. Such a microtubule will rotate as it travels across the surface (see inset).}
    \label{fig:schematic}
\end{figure}

In gliding assay experiments, the motor proteins are instead bound to a surface such that free microtubules placed over them are propelled over the surface in a manner reminiscent of ``crowd surfing''. The formerly  longitudinal component of the motor proteins' stepping is converted into a tangential force at each point along the microtubule, while the formerly azimuthal component is converted into a small transverse force which leads to rolling \cite{Meissner2024}. The net result is a screw-like propulsion, with rotation, of microtubules over the surface, as depicted schematically in  Fig.~\ref{fig:schematic}(a). Such screw-like propulsion has recently been proposed to be the source of chiral collective behaviors in ordered active matter states, both in gliding microtubules  \cite{athani2025gliding} and in colonies of some rod-shaped bacteria such as \textit{Myxococcus xanthus} \cite{banerjee2024active}. A complete picture of this type of chiral active matter requires us to understand the interplay between screw-like propulsion and filament flexibility at the single-filament level, which has not yet been explored, and which is our main focus in this work.

Though microtubules have a relatively large flexural rigidity \cite{Gittes1993, Venier1994, Hawkins2010} making them rigid with respect to thermal fluctuations, the active forces imposed by motor proteins walking along them are sufficiently large that they behave as \textit{semi-flexible} filaments in gliding assay experiments \cite{Sumino2012, Scharrel2014, Saito2017, Venier1994}.
As such, previous research has noted that individual microtubules in gliding assay experiments can exhibit multiple conformations, with straight configurations translating in straight paths while curved configurations follow smooth curved paths or loops. Figs.~\ref{fig:schematic}(b) and \ref{fig:schematic}(c) show schematic examples of such states. So far, this ``multi-stability'' has been attributed to either conformational switching \cite{Ziebert2015} or differential binding of motor proteins to the microtubules \cite{Pearce2018}; in practice, obstructions due to impurities likely also play a role \cite{ng2025active}. 

In this paper, we propose an alternative explanation of this multi-stability in shape and gliding trajectories, premised on the chiral screw-like motion of the microtubules. To this end, we develop a mathematical framework for studying individual one-dimensional elastic plane curves exposed to active forces and argue that this approach can provide insight into many features of actively driven filaments such as those seen in gliding assay experiments. Beyond explaining features of microtubule gliding assay experiments, the model is easily generalizable, capable of modeling far more general circumstances including stochastic forces such as thermal driving and forces generated by external potentials.

The structure of the paper is as follows. In Sections ~\ref{sec:Plane Curves} and \ref{sec:Elastic Filaments}, we develop the mathematical framework and apply it to active elastic filaments. Important particular stationary solutions are determined in Sec.~\ref{sec:Constant Solutions} and a linear stability analysis of them is carried out in Sec.~\ref{sec:Stability Analysis}. The nature of the time-dependent evolution towards these stationary solutions is considered in Sec.~\ref{sec:Time-dep Evolution}. Finally, simulations which validate many of our predictions are presented in Sec.~\Ref{sec:Simulations} and a discussion appears in Sec.~\ref{sec:Discussion}.

\section{Time-Dependent Evolution of Elastic Filaments}

\subsection{\label{sec:Plane Curves}Over-Damped Plane Curves}

We begin by recalling some features of plane curves, i.e.~one-dimensional curves embedded in two-dimensional space \cite{DoCarmoBook}. Any such curve can be parametrized by a function $\vec{r}(s):(0,L)\rightarrow\mathbb{R}^2$. The choice of parametrization is not unique; however, a natural choice can always be imposed by parametrizing by arc length (p.a.l.). Assuming the curve is differentiable, this amounts to imposing $|\vec{r}\,'(s)| = 1$ for all $s\in(0,L)$ in which case $L$ denotes the total length of the curve. If a curve is not p.a.l., uniform ``motion'' in $s$ does not necessarily correspond to uniform ``motion'' along the curve, i.e.~the curve is ``stretched'' or ``compressed'' relative to reference interval $(0,L)$.

Whether a curve is p.a.l.~or not, its parametrization's first derivative 
$\vec{r}\,'(s)$ is simply the tangent vector at each point $\vec{r}(s)$ and its magnitude $|\vec{r}\,'(s)|$ can be thought of as the ``stretching factor'' or ``metric'' at that point. To describe the orientation of the curve at each point, it is sufficient to consider the unit tangent vector 
\mbox{$\hat{T}(s) = \vec{r}\,'(s)/|\vec{r}\,'(s)|$}. For curves which are p.a.l., $\hat{T}(s) = \vec{r}\,'(s)$. Importantly, unlike the tangent vector $\vec{r}\,'(s)$, the unit tangent vector $\hat{T}(s)$ at every point along the curve is unchanged upon reparametrization of the curve and is thus an intrinsic property of the curve.

For plane curves, the curvature at each point is simply the rate of change of the unit tangent vector as one moves along the curve. Specifically, let $\hat{N}(s)$ denote the normal vector at each point, obtained by rotating $\hat{T}(s)$ $90^\circ$ anticlockwise. Since $\hat{T}(s) \cdot \hat{T}(s) = 1$, differentiation yields
$2\hat{T}(s) \cdot \hat{T}'(s) = 0$, i.e.~$\hat{T}'(s)$ is perpendicular to $\hat{T}(s)$ and thus parallel to 
$\hat{N}(s)$. Accordingly, it is natural to define the curvature $\mathsf{k}(s)$ as the magnitude of the derivative of the unit tangent vector 
$\hat{T}'(s) = \mathsf{k}(s)\hat{N}(s)$. Though this definition of  curvature is not intrinsic to the curve as it is not invariant under reparametrizations, an intrinsic curvature is given by rescaling by the magnitude of the tangent vector: $k(s) = \mathsf{k}(s)/|\vec{r}\,'(s)|$.

With this background and notation in hand, we now turn to the question of how to describe the time evolution of a plane curve's intrinsic properties. We imagine a time varying plane curve $\vec{r}(s)$, not necessarily p.a.l., subject to a net force per unit length $\vec{F}$. In the over-damped limit, inertia can be neglected and the curve's time evolution will be given by the differential equation
\begin{equation}
    \mu\frac{\partial\vec{r}(s)}{\partial t} = \vec{F} \,.
\label{eq:overdamped evolution}
\end{equation}
Here, $\mu$ denotes the (possibly direction dependent) damping coefficient per unit length of the curve, however without loss of generality, we can always rescale the components of $\vec{F}$ such that $\mu = 1$, and we will do so in the remainder of this paper.  Similarly, though $\vec{r}(s)$ and the various quantities derived from it are time dependent, for the sake of conciseness, we will tend not to explicitly denote its time dependence $t$. When it is helpful to denote the time explicitly, we will do so with a subscript, e.g.~$\vec{r}_t(s)$. Primes on quantities will unambiguously continue to denote derivatives along the curve, i.e.~with respect to $s$.

Differentiating $|\vec{r}\,'(s)|$, $\hat{T}(s)$ and 
$\mathsf{k}(s)$ with respect to time, we ultimately obtain the coupled equations
\begin{align}
    \frac{\partial |\vec{r}\,'(s)|}{\partial t} & = \vec{F}'\cdot\hat{T}(s) \,, \label{eq:stretch evolution} \\ 
    \frac{\partial\hat{T}(s)}{\partial t} & = \frac{\vec{F}'\cdot\hat{N}(s)}{|\vec{r}\,'(s)|}\hat{N}(s) \,, \label{eq:orientation evolution} \\ 
    \frac{\partial\mathsf{k}(s)}{\partial t} & = \left[\frac{\vec{F}'\cdot\hat{N}(s)}{|\vec{r}\,'(s)|}\right]' \,, \label{eq:curvature evolution}
\end{align}
where we have used Eq.~(\ref{eq:overdamped evolution}) to simplify these expressions. Whereas Eq.~(\ref{eq:overdamped evolution}) describes the evolution of the curve's position with time, these equations describe the evolution of its \textit{shape}. In particular, Eq.~(\ref{eq:stretch evolution}) states that the tangential component of the variation of the force along the curve determines how much it will stretch, while Eqs.~(\ref{eq:orientation evolution}) and (\ref{eq:curvature evolution}) state that the scaled normal component of the variation of the force determines how much the curve will rotate and bend. Since the systems we are interested in describing have forces that depend on the curve's intrinsic properties rather than its position, it will be much more natural to treat Eqs.~(\ref{eq:stretch evolution})-(\ref{eq:curvature evolution}) rather than Eq.~(\ref{eq:overdamped evolution}). Similar equations and their corresponding theory have been developed previously in
\cite{Nakayama1992, Nakayama1993}.

To build some intuition for Eqs.~(\ref{eq:stretch evolution}) through (\ref{eq:curvature evolution}), it is helpful to consider the uniformly curved filament conformation of Fig.~\added{\ref{fig:schematic}}(c). In order for such a filament to follow a circular trajectory with radius matching its conformation, the net force on each point must be everywhere tangential at all times. This is due to the overdamped dynamics, in contrast to an inertial scenario where a centripetal force would be required. If $\vec{F} = F_T \hat T$  has a constant tangential component $F_T$,  then  the right-hand side of Eq.~(\ref{eq:orientation evolution}) reduces to $F_T k \hat N(s)$. By assuming that the trajectory's radius of curvature $R$ matches that of the conformation $1/k$, we assume an angular rotation rate $ v / R = F_T k$, which shows the validity of Eq.~(\ref{eq:orientation evolution}) in this special case. We see that curvature of the conformation generates rotation of the tangent even when the net force has a uniform tangential component, and that no normal component of force is required to generate uniform circular motion.

Similarly, it may be considered counter-intuitive that Eqs.~(\ref{eq:stretch evolution}) and (\ref{eq:curvature evolution}) depend on the projection of the \textit{derivative} of the force, $\vec{F}'$, onto the tangential and normal directions $\hat{T}$ and $\hat{N}$, rather than the force $\vec {F}$ itself, as one might expect the tangential component of the force, $\vec{F}\cdot\hat{T}$, to induce stretching and the normal component, $\vec{F}\cdot\hat{N}$, to induce rotation and bending. To develop intuition for why this is not the case, one can consider the curve's response to uniform constant forces. For instance, suppose we begin with a straight conformation which experiences at each point a constant magnitude force in a purely tangential direction. This force will propel each point in an identical fashion and thus any pair of points will remain the same distance from each other. Accordingly, the curve's stretching factor $|\vec{r}\,'(s)|$ at each point will remain constant in time, precisely as predicted by Eq.~(\ref{eq:stretch evolution}). Similarly, if each point of a straight conformation experiences a constant magnitude force in a purely normal direction, each point will again be propelled in an identical fashion and thus the curve will simply translate through space with no rotation or bending, as predicted by Eqs.~(\ref{eq:orientation evolution}) and (\ref{eq:curvature evolution}). \removed{This observation embodies the fact that the total torque of a constant normal force along the entire length of the curve is zero.} As such, it is changes in the force, captured by $\vec{F}'$, which induce stretching, bending and rotation of a straight curve. 

While this argument only applies to an initially straight conformation, similar reasoning can be applied to curved conformations. For instance, consider again the uniformly curved state but now with each point experiencing a constant magnitude force $\vec F=F_N\hat N$ in a purely normal direction, say pointing towards the circle's center. In this case, the force will propel every point identically towards the center of the circle and thus the curve will become compressed without rotation\removed{ or a change in scaled curvature $\mathsf{k}(s)$ at any point, again} as predicted by Eq.~(\ref{eq:orientation evolution})\removed{ and (\ref{eq:curvature evolution})}. 
\added{As the radius $R$ of the arc shrinks, the curvature $k(s) = 1/R$ will diverge and thus the scaled curvature $\mathsf k(s) = |\vec r \,'(s)|k(s)$ remains constant, as predicted by Eq.~(\ref{eq:curvature evolution}).}
This remains the outcome even if a nonzero tangential component is added to the force, except that the compressed filament may now be rotated and translated. The filament compression is captured by Eq.~(\ref{eq:stretch evolution}) through the fact that in this case, $\vec{F}'= - F_N \mathsf{k}(s)\hat T$ is tangential and constant in magnitude, making the right-hand side non-zero.

Before turning to specific forces, it can be checked that Eq.~(\ref{eq:orientation evolution}) has the formal solution
\begin{equation}
    \hat{T}(s)=R\left[\int_0^t d\tau\,
    \frac{\vec{F}'\cdot\hat{N}(s)}{|\vec{r}\,'(s)|}\right] \hat{T}_0(s) \,, \label{eq:orientation solution}
\end{equation}
where $\hat{T}_0(s)$ denotes the unit tangent vector of the curve at time $t = 0$ and $R[\theta]$ denotes the rotation operator by angle $\theta$. In principle, Eqs.~(\ref{eq:stretch evolution}) and (\ref{eq:curvature evolution}) can depend on $\hat{T}(s)$, in which case these equations can either be treated as three coupled partial differential equations or two coupled partial integro-differential equations. When the driving force $\vec{F}$ is translationally and rotationally invariant, however, Eqs.~(\ref{eq:stretch evolution}) and (\ref{eq:curvature evolution}) cannot depend on the arbitrary orientation of the curve and thus decouple from $\hat{T}(s)$. Since we will only consider such forces which are invariant with rigid motions of the filament, we will focus on studying the solutions to Eqs.~(\ref{eq:stretch evolution}) and (\ref{eq:curvature evolution}), relying on Eq.~(\ref{eq:orientation solution}) to provide explicit results for the curve orientation.

\subsection{\label{sec:Elastic Filaments}Active Elastic Filaments}

We can model active elastic filaments, such as microtubules in a gliding assay experiment, as one-dimensional plane curves subject to stretching, bending and active forces.

In the continuum limit, stretching and bending forces can respectively be derived from the Hamiltonians
\begin{align}
    H_S[\vec{r}(s)] & = \frac{1}{2} \kappa_{S} 
    \int_0^L ds\, (|\vec{r}\,'(s)| - 1)^2 \,, \\
    H_B[\vec{r}(s)] & = \frac{1}{2} \kappa_B 
    \int_0^L ds\, \mathsf{k}(s)^2 \,,\label{eq:Hb}
\end{align}
where $\kappa_S$ and $\kappa_B$ denote the stretching and bending rigidity of the filament, and $L$ denotes the filament's \textit{unstretched} length. In principle, the integrand of Eq.~\ref{eq:Hb} could more generally read $[\mathsf{k}(s)-\mathsf{k}_I(s)]^2$, where 
$\mathsf{k}_I(s)$ is the an intrinsic (equilibrium) curvature. As the primary goal of this paper is to investigate whether naturally straight filaments subjected to chiral active forces can maintain stable curved conformations, we will take  $\mathsf{k}_I(s)=0$ throughout. The behavior of intrinsically curved filaments with purely tangential active driving has been studied previously~\cite{Ziebert2015, Pearce2018, Kim2018}.

Also note that it is not the curvature $k(s)$ but rather the scaled curvature $\mathsf{k}(s) = |\vec{r}\,'(s)|k(s)$ as experienced from ``within the curve'' which contributes quadratically to the bending energy. Upon varying the configuration $\vec{r}(s)$ by a small amount $\delta\vec{r}(s)$, the variation in the Hamiltonians 
\mbox{$\delta H[\vec{r}(s)] = H[\vec{r}(s)+\delta\vec{r}(s)] - H[\vec{r}(s)]$} can be written at lowest order in 
$\delta\vec{r}(s)$ as
\begin{multline}
    \delta H_S = 
    \kappa_S\left[(|\vec{r}\,'(s)|-1)\hat{T}(s) \cdot\delta\vec{r}(s)\right]_{s=0}^L \\  
    - \kappa_S\int_0^L ds \, 
    \left[(|\vec{r}\,'(s)|-1)\hat{T}(s)\right]'\cdot \delta\vec{r}(s) \label{eq:delta H_S}
\end{multline}
and
\begin{multline}
    \delta H_B = 
    \kappa_B\left[\frac{\mathsf{k}(s)}{|\vec{r}\,'\left(s\right)|}\hat{N}(s) \cdot\delta\vec{r}\,'(s)\right]_{s=0}^L \\
    - \kappa_B\left[\frac{\mathsf{k}'(s)}{|\vec{r}\,'(s)|}\hat{N}(s) \cdot\delta\vec{r}\,(s) \right]_{s=0}^L \\
    + \kappa_B\int_0^L ds\,\left[\frac{\mathsf{k}'(s)}{|\vec{r}\,'(s)|}\hat{N}(s)\right]' \cdot\delta\vec{r}(s) \,. \label{eq:delta H_B}
\end{multline}

(See Appendix~\ref{sec:Appendix} for a concise derivation of these expressions.)

Accordingly, the stretching force $\vec{F}_S$ and bending force $\vec{F}_B$ experienced by interior points of the filament are simply given by the variational derivative
\begin{equation}
    \vec{F} = - \frac{\delta H[\vec{r}(s)]}{\delta \vec{r}(s)} \,,
\end{equation}
i.e. for $s\in(0,L)$
\begin{align}
    \vec{F}_S & = \kappa_S 
    \left[(|\vec{r}\,'(s)|-1)\hat{T}(s)\right]' \,, \label{eq:FS}\\
    \vec{F}_B & = -\kappa_B \left[\frac{\mathsf{k}'(s)}{|\vec{r}\,'(s)|}\hat{N}(s)\right]' \,.
    \label{eq:FB}
\end{align}
The stretching and bending forces experienced at the boundaries $s=0$ and $s=L$ are, however, more subtle. 
For instance, with the aid of the Dirac delta-function $\delta(s)$, the boundary term of Eq.~(\ref{eq:delta H_S}) can be written as
\begin{multline}
    \left[(|\vec{r}\,'(s)|-1)\hat{T}(s) \cdot\delta\vec{r}(s)\right]_{s=0}^L = \\
    \int_0^L ds\,(|\vec{r}\,'(s)|-1)\hat{T}(s)[\delta(s-L)-\delta(s)] \cdot\delta\vec{r}(s)\,.
\end{multline}
This implies that the end points experience forces of infinite magnitude but only at the isolated end points such that the energy of a smooth arbitrary deformation remains finite. Accordingly, for the stretching force, it is natural to write the forces felt by the end points as
\begin{align}
    \vec{F}_S(s\rightarrow0) & = \lim_{s\rightarrow0}\kappa_S(|\vec{r}\,'(s)|-1)\hat{T}(s) \delta(s) \,,\\
    \vec{F}_S(s\rightarrow L) & = -\lim_{s\rightarrow L} \kappa_S(|\vec{r}\,'(s)|-1)\hat{T}(s)\delta(s-L) \,.
    \label{eq:FS at L}
\end{align}
To gain an intuition for why the end points experience such different forces to the interior points, consider a uniformly stretched straight filament. Under such circumstances, each interior point is uniformly stressed and thus would prefer for the overall system to shrink but since the forces acting on each element cancel out, no individual element will move. In contrast, this cancellation of forces doesn't occur at the end points where matter only exists to one side and thus the end points will experience a large net force inwards. Accordingly, a uniformly stretched filament will indeed shrink but this is (at least initially) entirely driven by the end points.

Similar issues arise with the bending forces at the filament ends. Here, it is helpful to rewrite the boundary terms of Eq.~(\ref{eq:delta H_B}) as
\begin{multline}
    \delta H_{B} = \kappa_B\left[\left(\frac{\mathsf{k}(s)}{|\vec{r}\,'(s)|}\hat{N}(s) \cdot\delta\vec{r}(s)\right)'\right]_{s=0}^L \\ 
    - \kappa_B\left[\left\{ \left(\frac{\mathsf{k}(s)}{|\vec{r}\,'(s)|}\hat{N}(s)\right)' + \frac{\mathsf{k}'(s)}{|\vec{r}\,'(s)|}\hat{N}(s)\right\} \cdot\delta\vec{r}\,(s)\right]_{s=0}^L \\
    + \kappa_B\int_0^L ds\, \left[\frac{\mathsf{k}'(s)}{|\vec{r}\,'(s)|}\hat{N}(s)\right]'\cdot\delta\vec{r}(s) \,. \label{eq:Hb_rewritten}
\end{multline}
As with the stretching force, the second of these terms implies infinite bending forces at the end points which can again be described using Dirac delta functions as
\begin{align}
    &\vec{F}_B(s\rightarrow0) = \nonumber \\
    & \enspace - \lim_{s\rightarrow0} \kappa_B \left\{ \left(\frac{\mathsf{k}(s)}{|\vec{r}\,'(s)|}\hat{N}(s)\right)' + \frac{\mathsf{k}'(s)}{|\vec{r}\,'(s)|}\hat{N}(s)\right\}\delta(s) \,, \label{eq:FB at 0}\\
    &\vec{F}_B(s\rightarrow L) = \nonumber \\
    & \enspace \lim_{s\rightarrow L} \kappa_B  \left\{ \left(\frac{\mathsf{k}(s)}{|\vec{r}\,'(s)|}\hat{N}(s)\right)' + \frac{\mathsf{k}'(s)}{|\vec{r}\,'(s)|}\hat{N}(s)\right\}\delta(s-L) \,.
\end{align}
In contrast, the first term in Eq.~(\ref{eq:Hb_rewritten}) only contributes in proportion to the derivative of the Dirac delta function, $\delta'(s)$, at the two ends. Physically, this corresponds to a net moment experienced by the filament which can affect the orientation of the curve but not its shape. As such, this term cannot provide a contribution to Eqs.~(\ref{eq:stretch evolution}) and (\ref{eq:curvature evolution}).

In microtubule gliding assay experiments, such as those described in the introduction, the microtubule filaments are driven by  motor proteins such as kinesin and dynein, anchored to the surface. These kinesins bind to the microtubules and, \textit{in vivo}, travel from the microtubule's positive end to the negative end or vice versa. However, because of the anchoring of the motors,  the microtubules are instead propelled across the surface. 

If the density of motors is sufficiently high, the propulsion can be approximated as an active force of constant and uniform magnitude at each point. The direction of this active force is roughly tangential to the filament. Numerous measurements, however, have shown that kinesins do not travel purely longitudinally along microtubules but rather have a nonzero azimuthal component to their motion  \cite{Meissner2024}. When the kinesins are anchored, this has the effect of rotating the microtubules like screws, which introduces a corresponding normal component to the filament motion. Accordingly, we model the microtubule-kinesin interaction by an active force $\vec{F}_A$ which points at a small constant angle $\alpha$ away from tangent of the filament
\begin{equation}
    \vec{F}_A = v\left[\cos(\alpha)\hat{T}(s) + \sin(\alpha)\hat{N}(s)\right]\,. \label{eq:FA}
\end{equation}
Here, $v$ denotes the active speed of a freely propelled filament.

With the stretching, bending and active forces at hand, we are ready to calculate the net force $\vec{F}$ experienced by each point on the filament as
\begin{equation}
    \vec{F} = \vec{F}_S + \vec{F}_B + \vec{F}_A \,. \label{eq:F total}
\end{equation}
In Eqs.~(\ref{eq:stretch evolution}) through (\ref{eq:curvature evolution}), it is the tangential and normal components of the derivative of the force, \mbox{$\vec{F}'\cdot \hat{T}$} and \mbox{$\vec{F}'\cdot \hat{N}$}, which enter the equations. After some careful manipulation, these quantities can be explicitly calculated as
\begin{multline}
    \vec{F}'\cdot \hat{T} = 
    \kappa_S\left[|\vec{r}\,'(s)|'' - \left(|\vec{r}\,'(s)|-1\right)\mathsf{k}(s)^{2}\right] \\
    + \kappa_B\left[\mathsf{k}(s)\left(\frac{\mathsf{k}'(s)}{|\vec{r}\,'(s)|}\right)' + \left(\frac{\mathsf{k}(s) \mathsf{k}'(s)}{|\vec{r}\,'(s)|}\right)'\right] \\
    - v\sin(\alpha)\mathsf{k}(s)
\end{multline}
and
\begin{multline}
    \vec{F}'\cdot \hat{N} = 
    \kappa_S\left[2|\vec{r}\,'(s)|' \mathsf{k}(s) + \left(|\vec{r}\,'(s)|-1\right)\mathsf{k}'(s)\right] \\
    + \kappa_B\left[\frac{\mathsf{k}(s)^{2}\mathsf{k}'(s)}{|\vec{r}\,'(s)|} - \left(\frac{\mathsf{k}'(s)}{|\vec{r}\,'(s)|}\right)''\right] \\
    + v\cos(\alpha)\mathsf{k}(s) \,. \label{eq:dF dot N}
\end{multline}
Accordingly, we find that Eqs.~(\ref{eq:stretch evolution}) and (\ref{eq:curvature evolution}) respectively become
\begin{multline}
    \frac{\partial |\vec{r}\,'(s)|}{\partial t} = 
    \kappa_S\left[|\vec{r}\,'(s)|'' - \left(|\vec{r}\,'(s)|-1\right)\mathsf{k}(s)^{2}\right] \\
    + \kappa_B\left[\mathsf{k}(s)\left(\frac{\mathsf{k}'(s)}{|\vec{r}\,'(s)|}\right)' + \left(\frac{\mathsf{k}(s) \mathsf{k}'(s)}{|\vec{r}\,'(s)|}\right)'\right] \\
    - v\sin(\alpha)\mathsf{k}(s) \label{eq:stretch evolution 2}
\end{multline}
and
\begin{multline}
    \frac{\partial\mathsf{k}(s)}{\partial t} = 
    \kappa_S\left[\frac{2|\vec{r}\,'(s)|'\mathsf{k}(s)+\left(|\vec{r}\,'(s)|-1\right)\mathsf{k}'(s)}{|\vec{r}\,'(s)|}\right]' \\
    + \kappa_B\left[\frac{\mathsf{k}(s)^2\mathsf{k}'(s)}{|\vec{r}\,'(s)|^2} - \frac{1}{|\vec{r}\,'(s)|} \left(\frac{\mathsf{k}'(s)}{|\vec{r}\,'(s)|}\right)'' \right]' \\
    + v\cos(\alpha)\left[\frac{\mathsf{k}(s)}{|\vec{r}\,'(s)|} \right]' \,. \label{eq:curvature evolution 2}
\end{multline}

Eqs.~(\ref{eq:stretch evolution 2}) and (\ref{eq:curvature evolution 2}) constitute a pair of coupled nonlinear partial differential equations for the evolution of $|\vec{r}\,'(s)|$ and $\mathsf{k}(s)$. Accordingly, given an initial configuration $\vec{r}_0(s)$ and appropriate boundary conditions, these equations can be integrated to determine the shape of the filament at any future time.

Eq.~(\ref{eq:overdamped evolution}) has no explicit boundary conditions associated with it. Rather, its boundary conditions at each time $t$ are determined by the solution $\vec{r}_\tau$ itself up to time $t$. Specifically, denoting the force per unit length $\vec{F}$ of Eq.~(\ref{eq:overdamped evolution}) at point $s$ and time $t$ by a function $\vec{F}(s;\vec{r}_t)$ of $s$ and the configuration $\vec{r}_t$, the boundary conditions at $s = 0,L$ are
\begin{align}
    \vec{r}_t(s=0) &= \vec{r}_0(s=0) + 
    \int_0^td\tau \,
    \vec{F}(s=0;\vec{r}_{\tau}) \,, \\
        \vec{r}_t(s=L) &= \vec{r}_0(s=L) + 
    \int_0^td\tau \,
    \vec{F}(s=L;\vec{r}_{\tau}) \,.
\end{align}
As such, without a complete solution to Eq.~(\ref{eq:overdamped evolution}), the boundary conditions of Eqs.~(\ref{eq:stretch evolution 2}) and (\ref{eq:curvature evolution 2}) cannot be determined. Some progress can be made, however, towards analyzing these equations by considering the following line of argument.

As discussed previously, the end points of the filament experience forces proportional to Dirac delta-functions at their ends. While the filament itself can be out of equilibrium, we might expect that the end points will rapidly equilibrate due to these large forces. Accordingly, we might expect force-free boundary conditions to roughly hold at the filament ends
\begin{equation}
    \vec{F}(s\rightarrow 0,L) \approx 0 \,.
    \label{eq:Force free BC}
\end{equation}
Separating tangential and normal components, we obtain the following four boundary conditions
\begin{align}
    \left[\kappa_S(|\vec{r}\,'(s)|-1)+\kappa_B\frac{\mathsf{k}(s)^{2}}{|\vec{r}\,'(s)|}\right]_{s=0,L} & \approx 0 \,, \label{eq:BC1} \\
    \left[\left(\frac{\mathsf{k}(s)}{|\vec{r}\,'(s)|}\right)'+\frac{\mathsf{k}'(s)}{|\vec{r}\,'(s)|}\right]_{s=0,L} & \approx 0 \,. \label{eq:BC2}
\end{align}

Upon careful examination, Eqs.~(\ref{eq:stretch evolution 2}) and (\ref{eq:curvature evolution 2}) are second order in their derivatives of $|\vec{r}\,'(s)|$ and fourth order in their derivatives of $\mathsf{k}(s)$. Accordingly, the coupled system is sixth order in spatial derivatives and thus we require 6 boundary conditions on the end points to uniquely integrate an initial condition through time. With Eqs.~(\ref{eq:BC1}) and (\ref{eq:BC2}), we have proposed 4 such boundary conditions by assuming that the end points are always approximately equilibrated. The remaining 2 boundary conditions cannot be written explicitly; they are instead determined instantaneously by the configuration $\vec{r}_t(s)$ at each time $t$. As will become clear in the next section, it is this apparent indeterminacy of the boundary conditions that makes multi-stability feasible.

To simplify the analysis of Eqs.~(\ref{eq:stretch evolution 2}) and (\ref{eq:curvature evolution 2}), it will be convenient to nondimensionalize them. Natural length and time scales include the length $L$  of the unstretched filament and the time $\tau = L/v$ taken for the active force $\vec{F}_A$ to propel the filament a distance equal to its length. These natural scales motivate defining dimensionless variables
\begin{align}
    \bar{s} & = s/L \,, \label{eq:nondim s}\\
    \bar{t} & = t/\tau \,, \\
    u(\bar{s}) & = |\vec{r}\,'(s)| \,, \label{eq: nondim u}\\
    w(\bar{s}) & = L\mathsf{k}(s) \label{eq:nondim k}\,.
\end{align}
We also introduce  two dimensionless parameters 
\begin{align}
    g_S & = \frac{\kappa_S\tau}{L^2} = \frac{\kappa_S}{Lv} \label{eq: gs}\\
    g_B & = \frac{\kappa_B\tau}{L^4} = \frac{\kappa_B}{L^3v}
\end{align}
which characterize the relative importance of stretching and bending forces in the system.
In terms of these rescaled quantities, Eqs.~(\ref{eq:stretch evolution 2}) and (\ref{eq:curvature evolution 2}) become
\begin{multline}
    \frac{\partial u}{\partial\bar{t}} = 
    g_S\left[u''-(u-1)w^2\right] \\
    + g_B\left[w\left(\frac{w'}{u}\right)' + \left(\frac{ww'}{u}\right)'\right] - \sin(\alpha)w \label{eq:u equation}
\end{multline}
and
\begin{multline}
    \frac{\partial w}{\partial \bar{t}} = 
    g_S\left[\frac{2wu'+(u-1)w'}{u}\right]' \\
    + g_B\left[\frac{w^2w'}{u^2} - 
    \frac{1}{u}\left(\frac{w'}{u}\right)''\right]' + \cos(\alpha)\left(\frac{w}{u}\right)' \,, \label{eq:w equation}
\end{multline}
with Eqs.~(\ref{eq:BC1}) and (\ref{eq:BC2}) becoming the boundary conditions
\begin{align}
    \left[g_S(u-1)u + g_Bw^2\right]_{\bar{s}=0,1} & \approx  0 \,, \label{eq:BC1 non dim} \\
    \left[2uw' - wu'\right]_{\bar{s}=0,1} & \approx 0 \label{eq:BC2 non dim} \,.
\end{align}

For microtubule gliding assay experiments, the orders of magnitude of $g_S$ and $g_B$ can be determined as follows. First, in experiments, microtubules have lengths of roughly $L\sim10\textrm{ µm}$ and travel with velocities around $v\sim0.5\textrm{ µm/s}$ \cite{Venier1994, Sumino2012, Scharrel2014, Saito2017}. From the fluctuation-dissipation relationship, their damping coefficient $\mu$, defined in Eq.~(\ref{eq:overdamped evolution}), can be assessed to be $\mu=\gamma/L\sim4\times10^{-3}\mathrm{\,kg \cdot m^{-1} \cdot s^{-1}}$,  where $\gamma\sim4\times10^{-8}\mathrm{\,kg \cdot s^{-1}}$ denotes the filament's coefficient of friction \cite{Memarian2021, Athani2024}. Now the stretching modulus, $\kappa_S$, is simply the product of the filament's Young's modulus, $E$, with its cross-sectional area, $A$, normalized by the damping coefficient $\mu$. For microtubules, experimental measurements give $E\sim 10^9 \mathrm{\,Pa}$ and $A\sim 200 \mathrm{\,nm^2}$ \cite{Gittes1993}, so $\kappa_S\sim 6\times10^{-5}\mathrm{\,m^2 \cdot s^{-1}}$. Finally, the bending rigidity of microtubules has been experimentally measured to be around $10^{-23}\mathrm{\,N \cdot m^{2}}$ \cite{Gittes1993, Venier1994, Hawkins2010} thus after normalization by the damping coefficient $\mu$, we assess the bending modulus $\kappa_B \sim 2.5\times10^{-21}\mathrm{\,m^4 \cdot s^{-1}}$. Consequently, for microtubule gliding assay experiments, we find that $g_S\sim10^7$ while $g_B\sim5$, i.e. $g_S$ and $g_B$ are separated by six to seven orders of magnitude. Physically, this corresponds to microtubules being extremely inextensible while also being semi-flexible at the energy scale determined by the active forces. 

Along with $g_S$ and $g_B$, the system is characterized by one additional dimensionless parameter, namely the angle, $\alpha$, at which the active force acts relative to the filament tangent. The magnitude of this angle can be determined from the pitch of the helical motion of the microtubules, which has been measured to be around 1~µm \cite{Meissner2024}. With a measured circumference of roughly 0.1~µm, we find the off-tangential angle to be $\alpha\sim\arctan(0.1\textrm{ µm}/1\textrm{ µm})\sim 0.1\textrm{ radians}$. Accordingly, $\alpha$ is small but not vanishingly small.
%Along with $g_S$ and $g_B$, the system is characterized by one additional dimensionless parameter, namely the angle, $\alpha$, at which the active force acts relative to the filament tangent. The magnitude of this angle can be determined from the pitch of the helical motion of the microtubules, which has been measured to be around 1~µm \cite{Meissner2024}. With a measured circumference of roughly 0.1~µm, we find the off-tangential angle to be $\alpha\sim\arctan(0.1\textrm{ µm}/1\textrm{ µm})\sim 0.1\textrm{ radians}$. Accordingly, $\alpha$ is small but not vanishingly small.

Before moving on to the analysis of these equations, it is worth identifying and discussing the various forces we have neglected. First, we have not incorporated any forces related to self-intersection, i.e.~the filaments are completely transparent with respect to themselves. In gliding assay experiments, filaments are indeed often able to ``intersect'' each other or more precisely, detach from the surface and pass over each other. This however has an energy cost and crossing segments undoubtedly apply additional forces to each other. Since self-intersection of a curve is fundamentally encoded in the position vector $\vec{r}(s)$ rather than its intrinsic properties, the formalism we have developed in this paper is not readily applicable to systems in which self-intersection is important. This will be a limitation to bear in mind when the solutions to our equations describe self-intersecting curves.

Second, we have not incorporated thermal coupling forces between the filaments and their environment. Such forces would typically be modeled by stochastic thermal forces $\vec{F}_T$ with mean zero and variance
\begin{equation}
    \left< F_{T,i}(s_1)F_{T,j}(s_2) \right> = 2D\delta_{ij}\delta(s_1 - s_2) \,,
\end{equation}
where $F_{T,i}$ denotes component $i$ of the force (either tangential or normal to the filament), $\delta_{ij}$ denotes the Kronecker delta symbol, $\delta(s)$ denotes the Dirac delta-function and $D$ denotes the magnitude of the thermal autocorrelations. While such forces can readily be incorporated with our formalism, they convert our equations from highly nonlinear coupled partial differential equations (PDEs), to highly nonlinear coupled stochastic differential equations (SDEs) thereby complicating their analysis substantially. Accordingly, we leave thermal considerations for future work.

Similarly, the active force we have defined above assumes a uniform density of motors across the surface such that the magnitude of the active force can be treated as a constant $v$. In physical experiments, the motor density will necessarily vary across the surface and thus should be described by spatially varying quenched noise. As the filament travels over this quenched landscape, it will experience an active force whose magnitude fluctuates and thus in principle, the active force should also be modeled stochastically. As with the thermal forces, accounting for such stochasticity complicates the analysis substantially and thus we neglect such considerations here.

\subsection{\label{sec:Constant Solutions}Constant Curvature Stationary Solutions}

Eqs.~(\ref{eq:u equation}) and (\ref{eq:w equation}) determine the time evolution of the shape of a flexible filament subject to stretching, bending and off-tangential active forces. Since such a system is driven out of equilibrium by the active force, there is no guarantee that its long-term behavior will be stationary. Accordingly, it is of particular interest to determine whether our equations support any stationary solutions and if so, whether they are stable relative to small perturbations.

Arguably, the simplest possible filament configurations are ones with constant curvature and thus we start by seeking out stationary solutions with constant curvature $\mathsf{k}(s) = \mathsf{k}_0$ or in dimensionless terms, stationary solutions $(u,w)$ for which $w(\bar{s}) = w_0$ is constant. Upon substitution into Eqs.~(\ref{eq:u equation})--(\ref{eq:BC2 non dim}), we obtain the coupled ordinary differential equations (ODEs)
\begin{align}
    0 = g_S\left[ u'' - (u-1)w_0^2 \right] - \sin(\alpha)w_0 \,, \label{eq:constant eq 1}\\
    0 = \left[2g_S\left(\frac{u'}{u}\right)' + \cos(\alpha)\left(\frac{1}{u}\right)'\right]w_0 \,, \label{eq:constant eq 2}
\end{align}
with boundary conditions
\begin{align}
    \left[g_S(u-1)u + g_Bw_0^2\right]_{\bar{s}=0,1} & \approx 0 \,, \label{eq:constant BC 1}\\
    \left[w_0u'\right]_{\bar{s}=0,1} & \approx 0 \label{eq:constant BC 2} \,.
\end{align}

We can identify two situations. If $w_0 =0 $, i.e.~the filament is uncurved, then Eqs.~(\ref{eq:constant eq 2}) and (\ref{eq:constant BC 2}) are satisfied identically while Eq.~(\ref{eq:constant eq 1}) reduces to the second order ODE 
\begin{equation}
    u'' = 0 \,,
\end{equation}
with boundary conditions
\begin{equation}
    u(0) \approx u(1) \approx 1 \,.
\end{equation}
This ODE has unique solution
\begin{equation}
    u(\bar{s}) = 1 \,.
\end{equation}
Thus we find that
\begin{equation}
    u = 1,\qquad w = 0
\end{equation}
constitutes a stationary, constant-curvature solution. In fact, this solution simply describes a straight unstretched filament translating at an angle $\alpha$ relative to its orientation.

The more interesting situation is the one in which \mbox{$w_0 \ne 0 $}. In this case, both Eqs.~(\ref{eq:constant eq 1}) and (\ref{eq:constant eq 2}) constitute second order ODEs for $u$ while Eqs.~(\ref{eq:constant BC 1}) and Eqs.~(\ref{eq:constant BC 2}) constitute four boundary conditions. This situation is highly over-determined; thus, unless some spectacular cancellation occurs, the equations will have no solution. To achieve this cancellation, we suppose the existence of a constant solution $u = u_0$, in which case Eqs.~(\ref{eq:constant eq 2}) and (\ref{eq:constant BC 2}) are again satisfied identically while Eqs.~(\ref{eq:constant eq 1}) and (\ref{eq:constant BC 1}) reduce to
\begin{align}
    g_S(u_0 - 1)w_0 & = - \sin(\alpha) \,, 
    \label{eq:u0-w0 relationship 1} \\ 
    g_S(u_0 - 1)u_0 & \approx - g_B w_0^2 \,. 
    \label{eq:u0-w0 relationship 2}
\end{align}
If $\alpha = 0$, then $w_0 = 0$ which we have already seen constitutes the uncurved trivial solution. Accordingly, we will henceforth assume, without loss of generality, that $\alpha > 0$; solutions for $\alpha <0$ can be obtained from these by mirror reflection. Then these equations can be simplified to give
\begin{align}
    u_0 & = \frac{g_B}{\sin(\alpha)} w_0^3 \,, \label{eq:u0 from w0} \\
    \frac{g_B}{\sin(\alpha)} w_0^4 & = 
    w_0 - \frac{\sin(\alpha)}{g_S} \label{eq:polynomial} \,.
\end{align}

\begin{figure}
    \centering
    \includegraphics[width=1.0\linewidth]{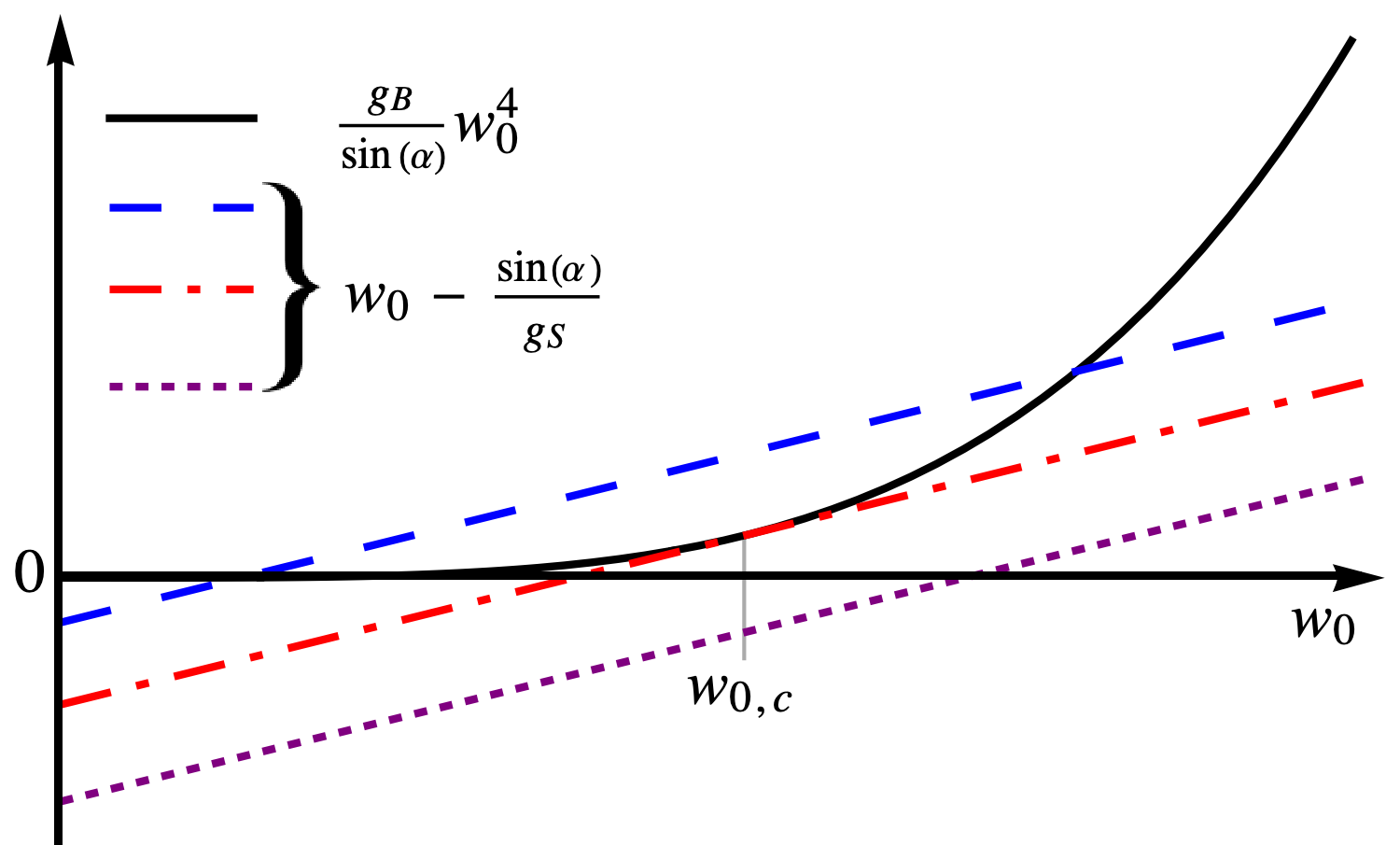}
    \caption{Schematic solutions to Eq.~(\ref{eq:polynomial}) for the curvature $w_0$ of uniformly curved solutions to the equations of motion. The points where the dashed lines intersect the solid curve (if any) constitute the roots of Eq.~(\ref{eq:polynomial}).
    There are two, one, or zero such roots if $g_b / \sin(\alpha)$ is, respectively, less than (blue line), equal to (red line) or greater than (purple line) $(3^3/4^4)(g_s / \sin(\alpha))^3$.
    %The three lines correspond to \mbox{$\frac{g_B}{\sin(\alpha)} < \frac{3^3}{4^4}\left(\frac{g_S}{\sin(\alpha)}\right)^3$} (blue dashed), \mbox{$\frac{g_B}{\sin(\alpha)} = \frac{3^3}{4^4}\left(\frac{g_S}{\sin(\alpha)}\right)^3$} (red dash-dotted), \mbox{$\frac{g_B}{\sin(\alpha)} > \frac{3^3}{4^4}\left(\frac{g_S}{\sin(\alpha)}\right)^3$} (purple dotted). 
    $w_{0,c}$ marks the value of $w_0$ for which Eq.~(\ref{eq:polynomial}) has only a single root.}
    \label{fig:polynomial solutions}
\end{figure}

For any given $w_0$, the first of these equations uniquely determines $u_0$ while the second of these equations states that $w_0$ is the root of a fourth order polynomial. Depending on the values of $g_B / \sin(\alpha)$ and $g_S / \sin(\alpha)$, this polynomial will have either 0, 1 or 2 real roots. Fig.~\ref{fig:polynomial solutions} shows a schematic of the left and right-hand sides of Eq.~(\ref{eq:polynomial}). For a fixed value of $g_B/\sin(\alpha)$ it can be seen that as $\sin(\alpha)/g_S$ is increased, the number of roots reduces from 2 to 1 to 0. At the critical value $w_{0,c}$ where only a single root exists, the derivatives of the left and right-hand sides of Eq.~(\ref{eq:polynomial}) must be equal such that
\begin{equation}
    w_{0,c} = \left(\frac{\sin\left(\alpha\right)}{4g_{B}}\right)^{1/3} \,.
\end{equation}
Substituting this value into Eq.(\ref{eq:polynomial}), we find the critical line
\begin{equation}
    \frac{g_B}{\sin(\alpha)} = 
    \frac{3^3}{4^4}\left(\frac{g_S}{\sin(\alpha)}\right)^3 \,. \label{eq:g_B-g_S critical line}
\end{equation}
Moving off this critical line, if \mbox{$g_B/\sin(\alpha) < (3^3/4^4) (g_S/\sin(\alpha))^3$}, Eq.~(\ref{eq:polynomial}) will have 2 real roots; otherwise, it will have no real roots.

\begin{figure}
    \centering
    \includegraphics[width=1.0\linewidth]{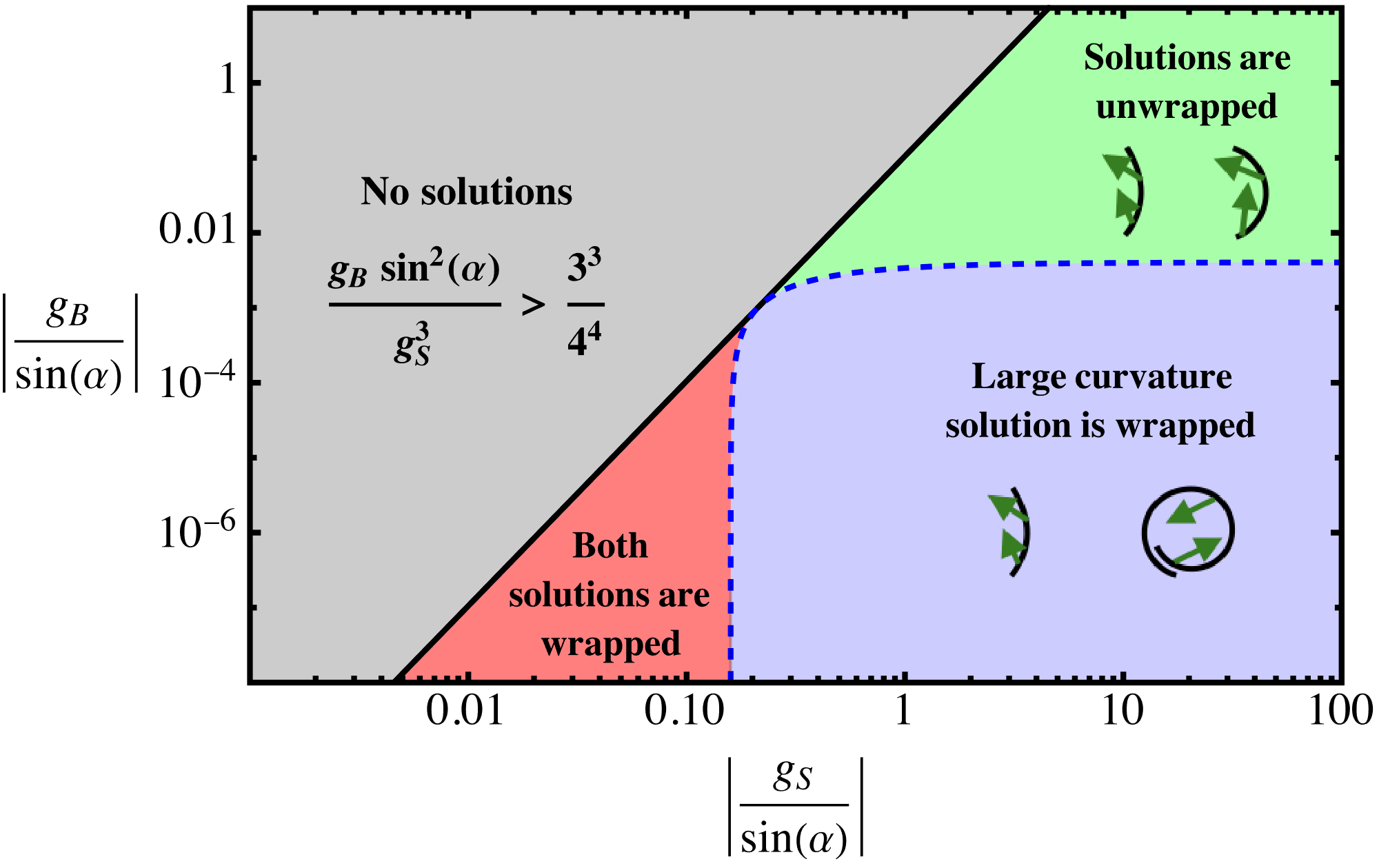}
    \caption{Phase space of non-trivial constant curvature solutions. The solid line corresponds to the critical line given by Eq.~(\ref{eq:g_B-g_S critical line}). On this line, Eq.~(\ref{eq:polynomial}) has exactly one real root, above it, there are no real roots, and beneath it, there are two real roots. The region with two real roots is then further divided by the blue dashed line, $w_0 = 2\pi$, which separates regions where both, one or neither solution fully wraps around and self-intersects.}
    \label{fig:phase space 1}
\end{figure}

We have thus found that our equations support non-trivial stationary, constant-curvature solutions of the form $(u_0,w_0)$. As these solutions constitute circular arcs, it is important to check whether they self-intersect. The length of any stretched curve is simply
\begin{equation}
    \mathsf{L} = \int_0^L ds\, |\vec{r}\,'(s)| = L\int_0^1 d\bar{s}\, u(\bar{s}) \,.
\end{equation}
Thus our circular arcs, which are uniformly stretched with 
$u = u_0$, will have length $\mathsf{L} = Lu_0$. At the same time, the circumference of a circle with constant curvature $k$ is just 
\begin{equation}
    \frac{2\pi}{k} = 
    \frac{2\pi|\vec{r}\,'(s)|}{\mathsf{k}(s)} = 
    \frac{2\pi L u(\bar{s})}{w(\bar{s})} = 
    \frac{2\pi L u_0}{w_0} \,.
\end{equation}
If the length of the arc exceeds this circumference, then the arc will overlap itself. Accordingly, we find that self-intersection will occur when $w_0 > 2\pi$. We refer to such states as ``wrapped''. Depending on the values of $g_S/\sin(\alpha)$ and $g_B/\sin(\alpha)$, it is possible for both, one or neither of the roots of Eq.~(\ref{eq:polynomial}) to describe a wrapped filament. Fig.~\ref{fig:phase space 1} shows a phase space of the possible situations.

As discussed in the previous section, our formalism does not handle self-intersection, so the situation where solutions wrap around onto themselves is not fully described by our model. It is reasonable to suppose, however, that a constant curvature filament for which self-intersection is prohibited might spiral under such circumstances. Indeed, spirals and whorls are occasionally observed in gliding assay experiments \cite{Weiss1991, Bourdieu1995, Kawamura2008, Kabir2012}.

The stationarity of these constant-curvature solutions means that they describe filaments whose shape remains unchanged, even as they translate and rotate through space, i.e. they behave as rigid bodies. Using Eqs.~(\ref{eq:orientation solution}) and (\ref{eq:dF dot N}), we can determine the rate at which they rotate. In terms of $u_0$ and $w_0$, we find that the unit tangent vector at each point evolves as
\begin{equation}
    \hat{T} = R\left[
    \frac{w_0\cos(\alpha)}{u_0}\bar{t}\right] \hat{T}_0 \,,
    \label{eq:T_hat sol const curvature}
\end{equation}
thus the dimensionless rate of rotation $\omega$ is given by
\begin{equation}
    \omega = \frac{w_0\cos(\alpha)}{u_0} = \frac{\sin(\alpha)\cos(\alpha)}{g_B w_0^2} \,,
    \label{eq:rotation rate}
\end{equation}
where Eq.~(\ref{eq:u0 from w0}) was used to derive the latter equality. 

While our primary goal in this paper is determining the possible shapes active elastic filaments can take, for the uniformly curved and compressed solution we have just derived, it is straightforward to also determine the filament's trajectory. Since the curvature $\mathsf{k}(s)$ and stretching factor $|\vec{r}\,'(s)|$ are constant, the total force experienced by any point in the filament, given by Eqs.~(\ref{eq:FS}), (\ref{eq:FB}), (\ref{eq:FA}) and (\ref{eq:F total}), is
\begin{multline}
    \vec{F}(s) = 
    \kappa_S (|\vec{r}\,'(s)|-1)\mathsf{k}(s)\hat{N}(s) \\
    +v\cos(\alpha)\hat{T}(s) +
    v\sin(\alpha)\hat{N}(s)\,.
    \label{eq:F total curved shape}
\end{multline}
The two terms in this force  proportional to $\hat N(s)$ cancel, as seen by expressing Eq.~(\ref{eq:u0-w0 relationship 1}) in dimensional terms using Eqs.~(\ref{eq: nondim u})--(\ref{eq: gs}). Therefore, the total force is strictly tangential for constant curvature solutions, and Eq.~(\ref{eq:overdamped evolution}) for the trajectory of the filament can be written as
\begin{equation}
    \frac{\partial\vec{r}(s)}{\partial \bar{t}} = L\cos(\alpha)\hat{T}(s)
    = L\cos(\alpha)R_{\omega \bar{t}}
    \hat{T}_0(s)\,,
\end{equation}
where we have used Eq.~(\ref{eq:T_hat sol const curvature}) to make the only time dependence explicit. Here, $R_\theta$ denotes the two dimensional rotation operator by angle $\theta$. It is now easy to integrate through with respect to time and obtain the trajectory
\begin{equation}
    \vec{r}(s)=\vec{r}_{0}(s)+\frac{u_{0}L}{w_{0}}(I-R_{\omega\bar{t}})\hat{N}_{0}(s) \,,
\end{equation}
where $\vec{r}_{0}(s)$ denotes the initial position of a point on the filament, $\hat{N}_0(s)$ denotes the initial normal vector at that point and $I$ denotes the identity matrix. If we choose to place the origin of our coordinate system at the center of the circular arc defined by our constant curvature solution, then we can equivalently write $\vec{r}_{0}(s) = 
-(Lu_0/w_0)\hat{N}_{0}(s)$ such that
\begin{equation}
    \vec{r}(s) = 
    R_{\omega\bar{t}}\vec{r}_0(s)
    = -\frac{Lu_0}{w_0} 
    R_{\omega\bar{t}}\hat{N}_0(s) \,.
\end{equation}
This expression makes it clear that each point in the filament rotates around the circular center of the filament arc, i.e.\ the filament is found to undergo uniform circular motion about the point equidistant from every point in the filament. While such behavior is to be expected from a rigid body experiencing only tangential forces, it is not a priori obvious that such motion would also emerge for a body out-of-equilibrium whose rigid shape is maintained by balancing internal stretching and bending forces with an external active force composed of both tangential and normal components.

\begin{figure}[t!]
    \centering
    \includegraphics[width=1.0\linewidth]{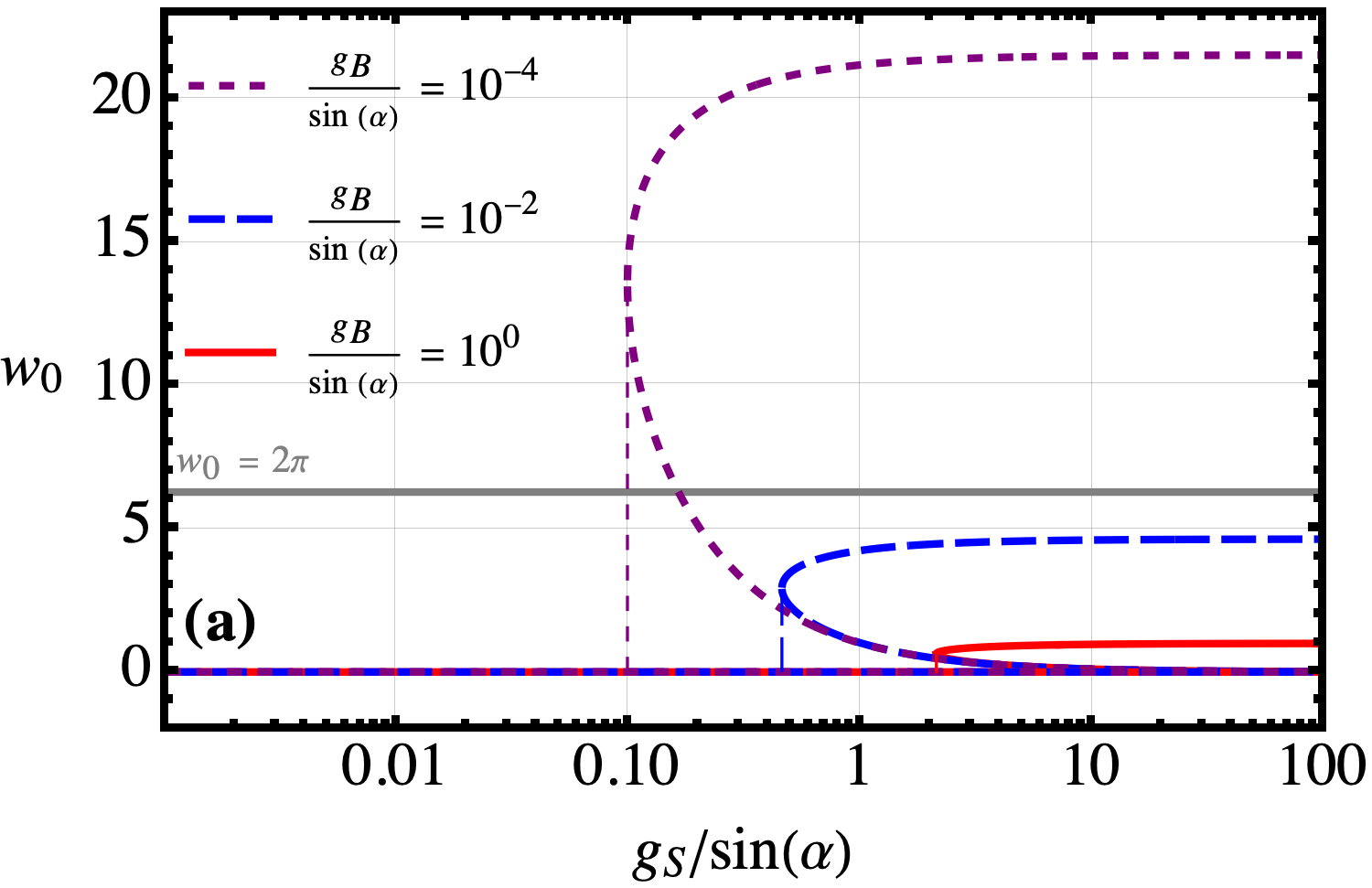}
    
    \vspace{0.1 cm}
    
    \includegraphics[width=1.0\linewidth]{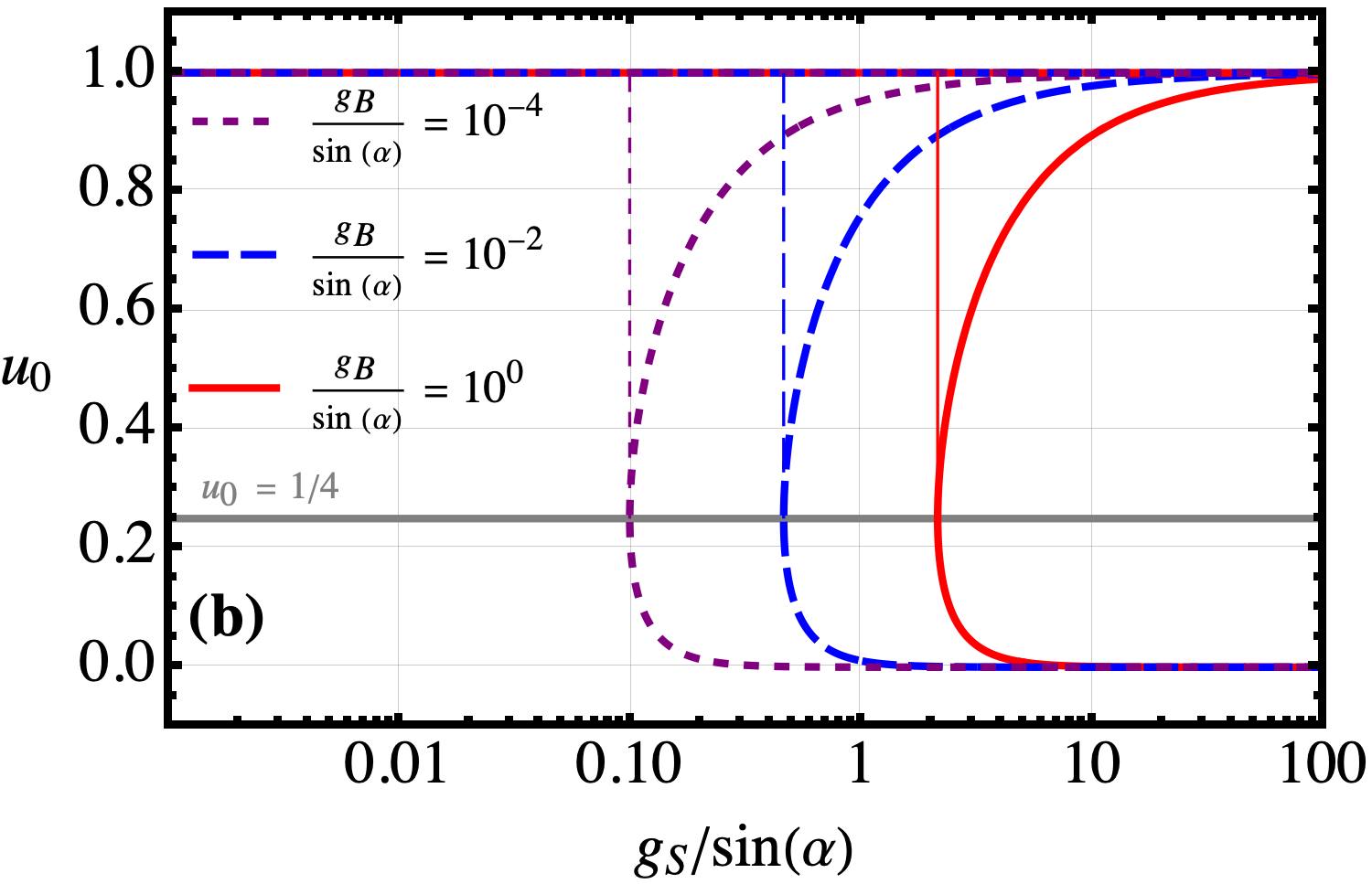}

    \vspace{0.1 cm}

    \includegraphics[width=1.0\linewidth]{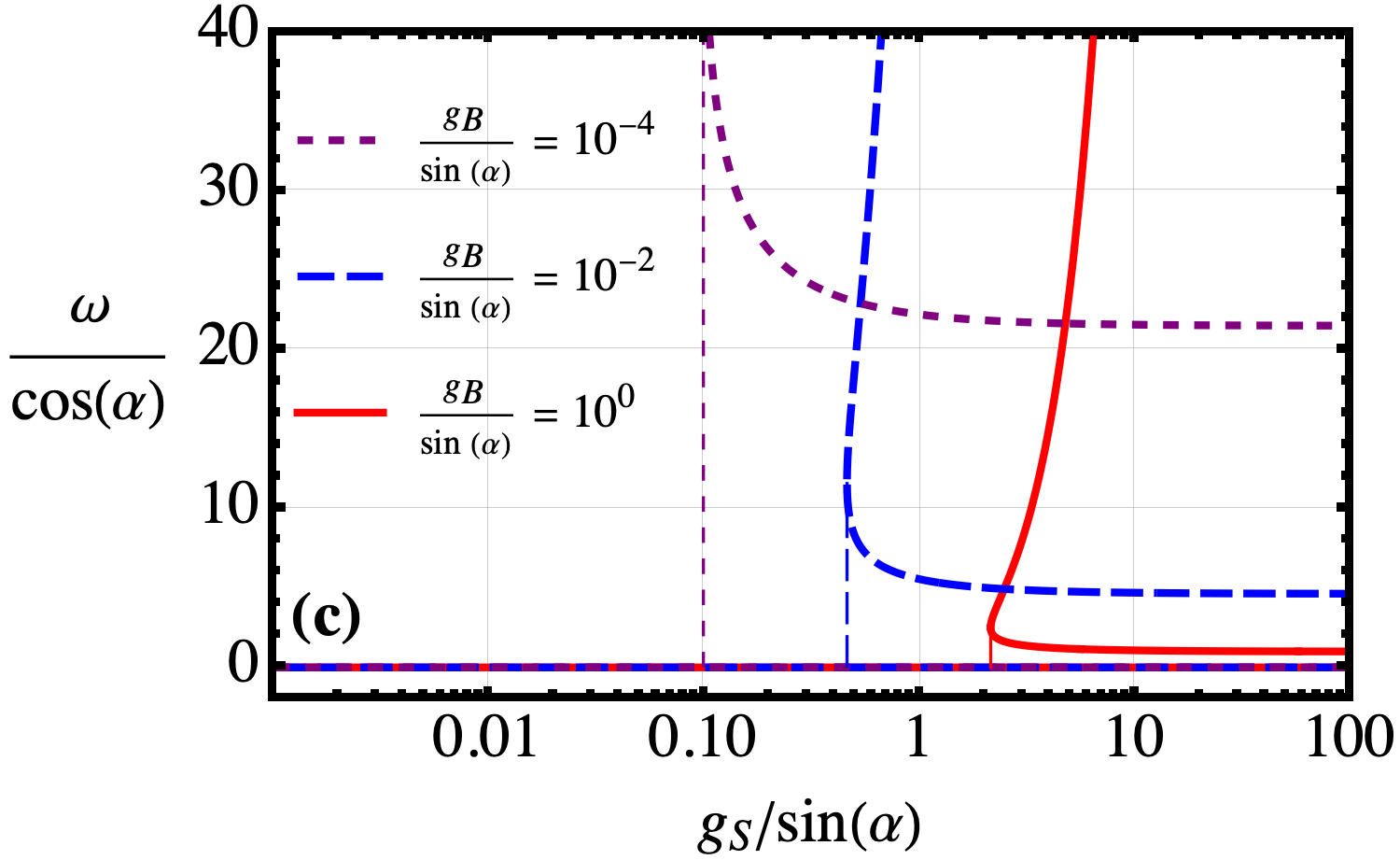}
    
    \caption{The theoretically predicted uniform solutions, as a function of  $g_S/\sin(\alpha)$, described in terms of their \textbf{(a)} uniform curvature $w_0$, \textbf{(b)} uniform stretching $u_0$, and \textbf{(c)} rotation rate $\omega/\cos(\alpha)$. The value of $g_B/\sin(\alpha)$ is given for each curve in the legends. The plots show that for small $g_S/\sin(\alpha)$, the only available constant-curvature solution is the uncurved ($w_0 = 0$), unstretched ($u_0 = 1$), non-rotating ($\omega = 0$) state. Above a critical value which depends on $g_B/\sin(\alpha)$, additional constant curvature solutions become possible.}
    \label{fig:w, u, omega vs gS}
\end{figure}

Fig.~\ref{fig:w, u, omega vs gS} shows how $w_0$, $u_0$ and $\omega/\cos(\alpha)$ vary as a function of $g_S/\sin(\alpha)$ for various choices of $g_B/\sin(\alpha)$. For each choice of $g_B/\sin(\alpha)$, these plots show that for small $g_S/\sin(\alpha)$, only the uncurved ($w_0 =0 $), unstretched ($u_0 = 1$), non-rotating ($\omega = 0$) solution exists. Then, once $g_S/\sin(\alpha)$ reaches a critical value, a subcritical ``blue sky'' bifurcation occurs,
also known as a subcritical saddle-node bifurcation, in which two new states suddenly become accessible ``out of the clear blue sky'' \cite{StrogatzBook},
 and curved, compressed, rotating solutions emerge. In particular, two branches occur, corresponding to ``large'' and ``small'' curvature states. For increasing $g_S/\sin(\alpha)$, the large curvature branch tends to a constant value (possibly larger than $2\pi$ and thus wrapped) while the small curvature branch tends to 0, i.e. an uncurved state. From Fig.~\ref{fig:w, u, omega vs gS}(b) [and Eq.~(\ref{eq:u0 from w0})], we see that this small curvature branch is also characterized by $u_0\rightarrow0$, i.e. the filament becomes compressed to a point. This is, of course,  completely unphysical, suggesting that the proposed force-free boundary condition cannot be physically realized. More precisely, for a highly compressed filament, we might expect even small perturbations from uniform compression to give rise to forces much larger than those experienced by the filament ends and thus the argument that the end forces should equilibrate effectively instantaneously cannot be justified. Similarly, Fig.~\ref{fig:w, u, omega vs gS}(c) shows that the rotation rate of the highly compressed filament explodes, which is again unphysical. 

Indeed, at the critical point where the 
bifurcation occurs, Fig.~\ref{fig:w, u, omega vs gS}(b) shows that $u_0$ becomes equal to exactly $1/4$ for \textit{any} choice of $g_B/\sin(\alpha)$, which is already a highly compressed state. This suggests that it is not just the small curvature branch which is physically unrealizable. Rather, the first part of the large curvature branch must also be unrealizable. Accordingly, though our equations state that two uniformly curved and compressed possible states should exist for sufficiently large stretching moduli, it would be surprising to find all such solutions physically realizable.

However, 
for the parameter regime relevant to microtubule gliding assays, one of our constant curvature solutions is physically realistic. (It coexists with a straight,  unstretched solution that is likewise physical.) In Sec.~\ref{sec:Elastic Filaments}, we assessed the magnitude of our dimensionless parameters for typical gliding assay experiments and found $g_S\sim10^7$, $g_B\sim5$ and $\alpha\sim0.1$. Accordingly, we examine the inextensible limit of $g_S\gg1$ for which simple approximations for the stretching factor, $u_0$, and curvature, $w_0$, can be obtained. To this end, we can define a scaled curvature
\begin{equation}
    \bar{w}_0=\left(\frac{g_B}{\sin(\alpha)}\right)^{1/3}w_0 \,,
\end{equation}
such that Eq.~(\ref{eq:polynomial}) can be written in terms of a single dimensionless parameter as
\begin{equation}
    \bar{w}_0^4=\bar{w}_0 - \left(\frac{g_B\sin^2(\alpha)}{g_S^3}\right)^{1/3} \,. \label{eq:scaled polynomial}
\end{equation}
In the limit of $g_S\rightarrow\infty$, this equation is easily solved and has solutions $\bar{w}_0 \rightarrow 0$ and $\bar{w}_0 \rightarrow 1$. Upon substitution into Eq.~(\ref{eq:u0 from w0}), we find the first of these solutions describes an uncurved filament with the unphysical property $u_0 \rightarrow 0$, meaning the end-to-end length of the filament shrinks to 0. We will therefore only focus on the solution $\bar{w}_0 \rightarrow 1$, for which $u_0 \rightarrow 1$, i.e. the filament is completely uncompressed, a very reasonable result for the unstretchable limit taken. 

Carrying out a perturbative expansion of Eq.~(\ref{eq:scaled polynomial}) around the solution $\bar{w}_0 = 1$, we find that
\begin{multline}
    \bar{w}_0 = 1 - \frac{1}{3}\left(\frac{g_B\sin^2(\alpha)}{g_S^3}\right)^{1/3} \\ 
    + O\left(\left[\frac{g_B\sin^2(\alpha)}{g_S^3}\right]^{2/3}\right) \,,
\end{multline}
from which we obtain
\begin{multline}
    w_0 = \left(\frac{\sin(\alpha)}{g_B}\right)^{1/3}\Bigg[ 1 - \frac{1}{3}\left(\frac{g_B\sin^2(\alpha)}{g_S^3}\right)^{1/3} \\ 
    + O\left(\left[\frac{g_B\sin^2(\alpha)}{g_S^3}\right]^{2/3}\right) \Bigg]
    \label{eq:w0 large g_S}
\end{multline}
and 
\begin{multline}
    u_0 = 1 - \left(\frac{g_B\sin^2(\alpha)}{g_S^3}\right)^{1/3} \\ 
    + O\left(\left[\frac{g_B\sin^2(\alpha)}{g_S^3}\right]^{2/3}\right) \,.
\end{multline}
This solution has rotation rate
\begin{multline}
   \omega = \cos(\alpha)\left(\frac{\sin(\alpha)}{g_B}\right)^{1/3}\Bigg[1+\frac{2}{3}\left(\frac{\sin^2(\alpha)g_B}{g_S^3}\right)^{1/3} \\ + O\left(\left[\frac{\sin^2(\alpha)g_B}{g_S^3}\right]^{2/3}\right)\Bigg] \,.
\end{multline}
For the particular values $g_S\sim10^7$, $g_B\sim5$ and $\alpha\sim0.1$, these expressions become
\begin{align}
    w_0 &\sim 0.3\times[1 - 10^{-8}]\,,\\
    u_0 &\sim 1 - 3\times10^{-8}\,,\\
    \omega &\sim 0.3\times[1 + 2\times10^{-8}]\,.
\end{align}
Upon comparison with available experimental videos of gliding assay experiments, these predictions are found to be the right order of magnitude. For instance, in Movie~S1 of \cite{Scharrel2014}, focusing on microtubules which follow curved trajectories, crude measurements give the ratio of the microtubules' lengths to their radii of curvature ($w_0$) as being around 0.2 and the ratio of their rotation rate to their velocity per length ($\omega$) as also being around 0.2.

\subsection{\label{sec:Stability Analysis}Linear Stability Analysis}

In the previous section, we found that the parameter values $g_S/\sin(\alpha)$ and $g_B/\sin(\alpha)$ determine which constant curvature solutions are viable stationary shapes for our active filament. In particular, we found that the trivial unstretched, uncurved shape is always a solution. Additionally, solutions with uniform, nonzero curvature can exist so long as $g_B/\sin(\alpha)$ is sufficiently small relative to $g_S/\sin(\alpha)$, i.e.~the filaments must be sufficiently flexible to support non-trivial constant curvature stationary shapes.

We now wish to determine which of these solutions are stable relative to small perturbations. Eqs.~(\ref{eq:u equation}) and (\ref{eq:w equation}) are highly nonlinear and thus a full treatment of their stability properties is extremely challenging. A linearized version, however, is relatively manageable. Let $\delta u$ and $\delta w$ denote arbitrary small perturbations and consider the state
\begin{align}
    u & = u_0 + \delta u \,,\\
    w & = w_0 + \delta w \,,
\end{align}
where $u_0$ and $w_0$ denote some stationary constant curvature solution to the equations. We seek to understand how this state evolves in time. Substituting into Eqs.~(\ref{eq:u equation}) and (\ref{eq:w equation}), we can expand up to lowest order in $\delta u$ and $\delta w$ to obtain
\begin{multline}
    \frac{\partial \delta u}{\partial \bar{t}} = 
    g_S[\delta u'' - 2(u_0 - 1)w_0\delta w - w_0^2\delta u] \\
    + 2g_B\frac{w_0}{u_0}\delta w'' - \sin(\alpha)\delta w
\end{multline}
and
\begin{multline}
    \frac{\partial \delta w}{\partial \bar{t}} = 
    g_S\left[ \frac{2w_0\delta u + (u_0 - 1)\delta w}{u_0} \right]'' \\
    + g_B\left[ \frac{w_0^2\delta w - \delta w''}{u_0^2} \right]'' 
    + \cos(\alpha)\left[\frac{u_0\delta w - w_0\delta u }{u_0^2} \right]' \,.
\end{multline}
Since these equations are linear by design, we can treat them with Fourier analysis. Expanding the perturbations in terms of their time dependent Fourier modes
\begin{align}
    \delta u &= \sum_n U_n(\bar{t}\,)e^{2\pi i n \bar{s}} \,,\\
    \delta w &= \sum_n W_n(\bar{t}\,)e^{2\pi i n \bar{s}} \,,
\end{align}
we can substitute these expressions into our linearized equations and equate coefficients to obtain equations for the time evolution of each mode. Factoring the modes $U_n$ and $W_n$, these equations can be concisely written in matrix form as
\begin{equation}
    \frac{\partial}{\partial \bar{t}} 
    \left(\begin{array}{c}U_{n}\\ W_{n}\end{array}\right) = - A_n \left(\begin{array}{c}U_{n}\\W_{n}\end{array}\right) \,,
\end{equation}
where $A_n$ is the $2\times2$ matrix
\begin{widetext}
\begin{equation}
    A_n =
    \left(\begin{array}{cc}
g_S\left(w_0^2+(2\pi n)^2\right) & 
2g_B\frac{w_0}{u_0}(2\pi n)^2 - \sin(\alpha) \\
2g_S\frac{w_0}{u_0}(2\pi n)^2 + i\frac{\cos(\alpha)w_0}{u_0^2}(2\pi n) & 
\frac{g_B}{u_0^2}(2\pi n)^4 - i\frac{\cos(\alpha)}{u_0}(2\pi n)
\end{array}\right) \,,
\end{equation}
\end{widetext}
and we have used Eqs.~(\ref{eq:u0-w0 relationship 1}) and (\ref{eq:u0-w0 relationship 2}) to simplify the components of $A_n$.

The modes $(U_n,W_n)$ will decay if the real parts of the eigenvalues of $A_n$ are positive; an arbitrary perturbation $(\delta u,\delta w)$ will only decay if \textit{all} modes $(U_n,W_n)$ decay. We begin by examining the uncurved, unstretched trivial solution, $(u_0,w_0) = (1,0)$ for which
\begin{equation}
    A_n = 
    \left(\begin{array}{cc}
g_S(2\pi n)^2 & \sin(\alpha) \\
0 & g_B(2\pi n)^4 - i\cos(\alpha)(2\pi n)
\end{array}\right) \,.
\end{equation}
The eigenvalues of such a matrix are simply its diagonal values
\begin{align}
    \lambda_{1,n} & = g_S(2\pi n)^2 \,, \\
    \lambda_{2,n} & = g_B(2\pi n)^4 - i\cos(\alpha)(2\pi n) \,.
\end{align}
Since both of these eigenvalues have positive real part for all $n$, we find that the uncurved, unstretched trivial solution, $(u_0,w_0) = (1,0)$ is linearly stable to small perturbations, independent of the parameters $g_S$, $g_B$ and $\alpha$. 

For non-trivial curved solutions $(u_0,w_0)$, matters are more complicated. While the full expressions for the eigenvalues of $A_n$ are long and provide little insight, their large $n$ behavior is
\begin{align}
\mathrm{Re}\left[\lambda_{1,n}\right] & = 
g_S\left[(2\pi n)^2 - 3w_0^2 \right] + O\left((2\pi n)^{-2}\right) \,, \\
\mathrm{Re}\left[\lambda_{2,n}\right] & = 
\frac{\sin^2(\alpha)}{w_0^6 g_B}(2\pi n)^4 + 4g_S w_0^2 + O\left((2\pi n)^{-2}\right) \,.
\end{align}
Thus, for sufficiently large $n$, both eigenvalues are guaranteed to have positive real part, i.e.~for any perturbation, the high frequency modes are guaranteed to decay. Accordingly, for each choice of $g_S$, $g_B$ and $\alpha$, only a finite number of small frequency modes need to checked to determine stability.

\begin{figure}
    \centering
    \includegraphics[width=1.0\linewidth]{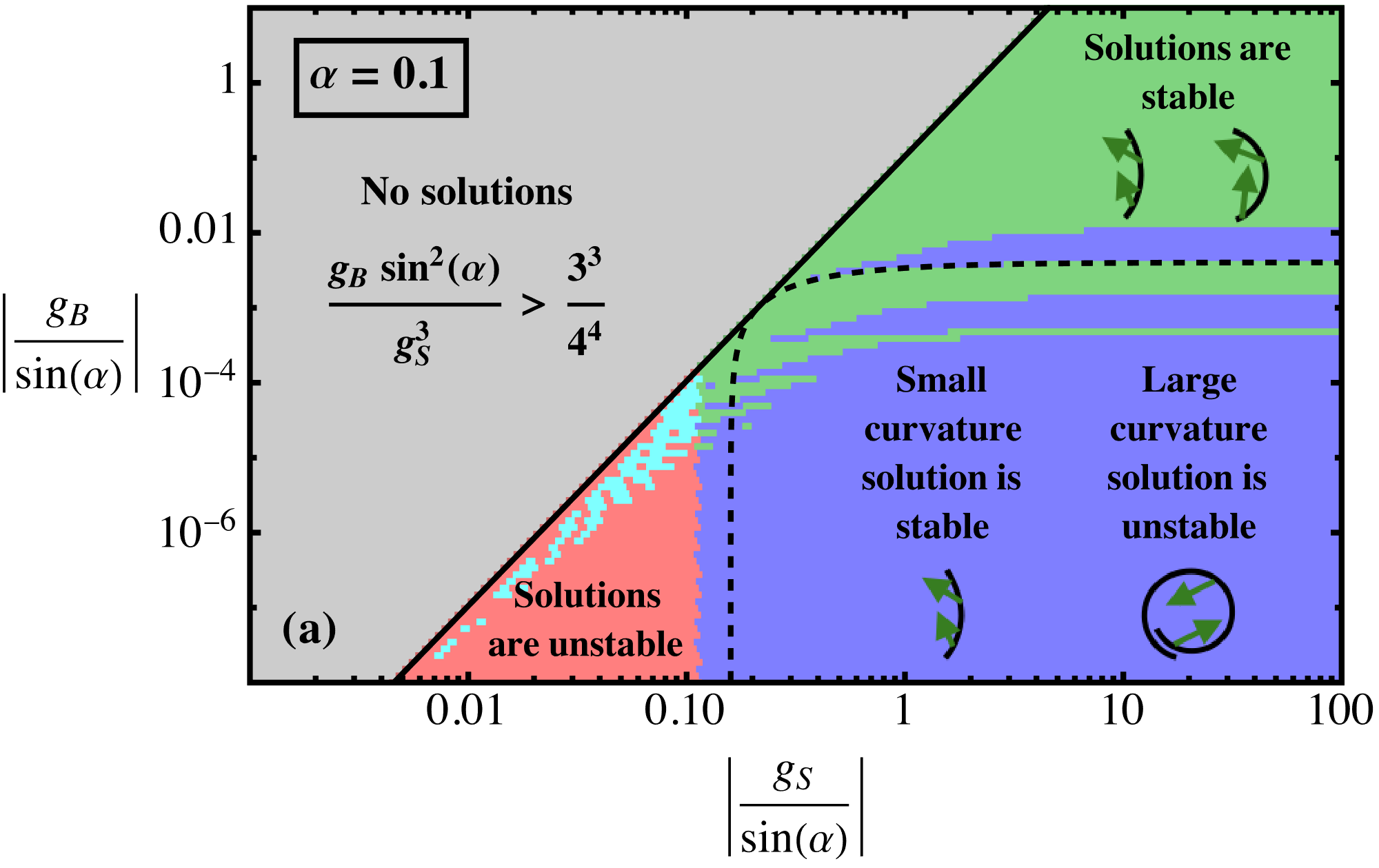}
    
    \vspace{0.4 cm}
    
    \includegraphics[width=1.0\linewidth]{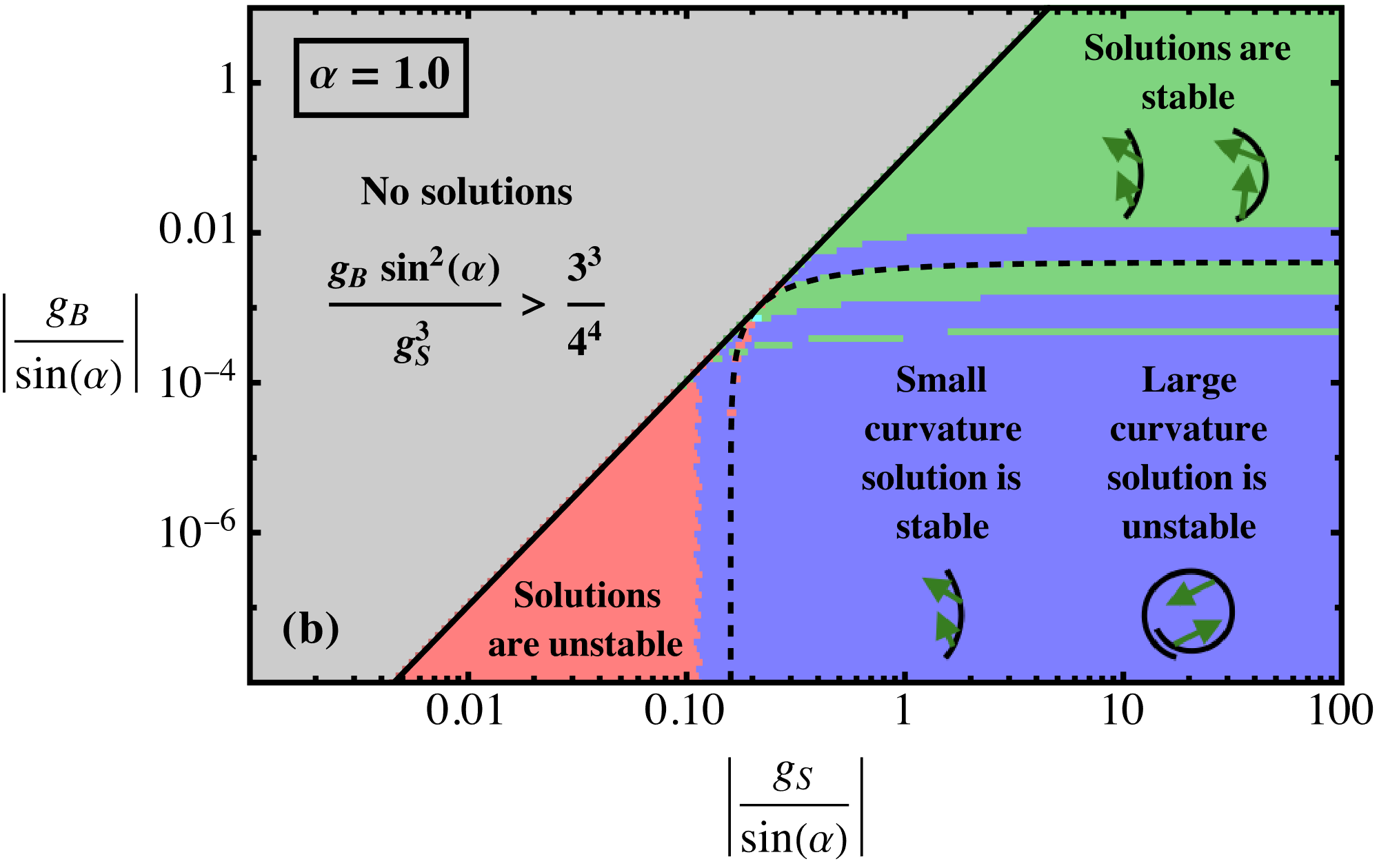}
    \caption{Linear stability of the curved solutions $(u_0,w_0)$ for \textbf{(a)} $\alpha = 0.1$, \textbf{(b)} $\alpha = 1.0$. The green regions show where both roots of Eq.~(\ref{eq:polynomial}) are stable and the red regions show where both roots are unstable. In the blue regions, the root with smaller curvature is stable while the larger-curvature root is unstable. The opposite is the case for the cyan regions. The dashed line is the same as that in Fig.~\ref{fig:phase space 1} and separates  the regions where one or both solutions self-intersect by  wrapping around.}
    \label{fig:phase space 2}
\end{figure}

Fig.~\ref{fig:phase space 2} shows the stability of each curved solution on the phase space $(g_S/\sin(\alpha),g_B/\sin(\alpha))$ for two different choices of $\alpha$. The green regions denote where both roots of Eq.~(\ref{eq:polynomial}) are stable while the red regions denote where both roots are unstable. In the blue and cyan regions, one of the roots is stable while the other is unstable. The dashed line is the same line that appears in Fig.~\ref{fig:phase space 1} and identifies where one or both of the roots result in a solution which wraps around and intersects itself. Curiously, we find that wrapping around is associated with the onset of instability, as the dashed line  mostly coincides with boundaries between green and blue regions or lies close to the boundary between red and blue regions. It is unclear why this should be the case. 

It is also interesting to observe that the value of $\alpha$ only makes a small difference to the stability phase space, its main effect being that of apparently stabilizing some ``wrapped around'' solutions. 

For both values of $\alpha$ examined in Fig.~\ref{fig:phase space 2}, an interesting banding feature emerges in the region separating the green and blue sections. This can be explained as follows. The functions $\mathrm{Re}[\lambda_{1,n}]$ and $\mathrm{Re}[\lambda_{2,n}]$ are continuous functions of $n\in\mathbb{R}$ though the stability criterion only requires that they be positive at all \textit{integer} values $n\in\mathbb{Z}$. This is because filaments of finite length only support discrete Fourier modes. Accordingly, these functions can have a negative minimum trough so long as the entire interval over which they are negative contains no integer. As $g_S$ and $g_B$ are varied, the location and width of this trough moves, sometimes including an integer value, sometimes not. The green bands within the blue region are those regions where at least one of the functions has a negative minimum but this minimum falls precisely between two integers and thus doesn't induce instability.

Though this stability analysis suggests that the small curvature solution is linearly stable so long as $g_S/\sin(\alpha)$ is sufficiently large, in the previous section, we argued that the small curvature solutions are not physically realizable at all as they correspond to highly compressed states. 
%Accordingly, little stock should be placed in the conclusions of this stability analysis regarding the small curvature solutions. 
Furthermore, near the 
bifurcation described by the solid black lines in Fig.~\ref{fig:phase space 2}, \textit{both} curved solutions have unphysically large compression. Accordingly, though the above stability analysis states that constant curvature solutions can be linearly stable right up to the critical line, we expect this particular conclusion to only be meaningful for points sufficiently far away from it.

\subsection{\label{sec:Time-dep Evolution}Time-Evolution of Filament Shape}

In Sec.~\ref{sec:Constant Solutions}, we determined the existence of stationary solutions to Eqs.~(\ref{eq:u equation}) and (\ref{eq:w equation}). The next natural question is whether we can understand the time evolution of filaments towards such states. Due to the highly nonlinear dynamics, this is generically difficult, but progress can be made by considering simplifying limits.

As discussed previously, microtubules are extremely inextensible and thus characterized by the limit $g_S\gg1$. Define the strain $\delta u(\bar{s})$ throughout the filament by
\begin{equation}
    u(\bar{s}) = 1 + \delta u(\bar{s}) \,.
\end{equation}
Then for $g_S\gg1$, we can suppose that $\delta u \sim O(1/g_S)$ and thus Eqs.~(\ref{eq:u equation}) and (\ref{eq:w equation}) can be written up to lowest order as
\begin{multline}
    \cancel{\frac{\partial\delta u}{\partial\bar{t}}} = 
    g_S[\delta u''-\delta uw^2] 
    + g_B[(ww')'+ww''] \\
    - \sin(\alpha)w + O(1/g_S) \,
    \label{eq:u equation limit}
\end{multline}
and
\begin{multline}
    \frac{\partial w}{\partial\bar{t}} = 
    g_S[2w\delta u'+w'\delta u]' 
    - g_B\left[w''-\frac{1}{3}w^3\right]'' \\ 
    + \cos(\alpha)w' + O(1/g_S) \,.
    \label{eq:w equation limit}
\end{multline}
In these equations, we have dropped terms that only contribute at order $O(1/g_S)$. Since $\delta u $ contributes at this order, its time derivative is inconsequential and this justifies its cancellation in Eq.~(\ref{eq:u equation limit}). Conceptually, we are asserting that the strain $\delta u$ evolves much faster than the curvature $w$ and this separation of time scales allows us to approximate Eqs.~(\ref{eq:u equation limit}) and (\ref{eq:w equation limit}) as a second order ODE coupled to a fourth order PDE. The boundary conditions for these equations are then determined by taking the large $g_S$ limit of Eqs.~(\ref{eq:BC1 non dim}) and (\ref{eq:BC2 non dim}) which ultimately gives
\begin{align}
    \delta u(\bar{s}=0,1) &\approx 
    - \frac{g_B}{g_S}w(\bar{s}=0,1)^2 \,, 
    \label{eq:BC 1 limit}\\
    w'(\bar{s}=0,1) &\approx 0 \,. \label{eq:BC 2 limit}
\end{align}
For any given curvature profile $w(\bar{s})$, the first of these boundary conditions can be used with Eq.~(\ref{eq:u equation limit}) to uniquely determine the strain profile $\delta u(\bar{s})$ throughout the filament. This can then be used in Eq.~(\ref{eq:w equation limit}) to evolve the curvature profile through time. As a fourth order PDE, the time evolution requires four boundary conditions, of which only two are provided by Eq.~(\ref{eq:BC 2 limit}). The remaining two are only defined implicitly from moment to moment by the filament configuration $\vec{r}_{\bar{t}}(\bar{s})$. Though lacking these boundary conditions places severe constraints on how much we can say about the time evolution of the filament, much can nevertheless be inferred.

In particular, let us consider the three terms on the right-hand side of Eq.~(\ref{eq:w equation limit}) separately. To understand the contribution of each term, we consider hypothetical scenarios in which one term dominates over the other two. Suppose only the last term contributes significantly, such that
\begin{equation}
    \frac{\partial w}{\partial\bar{t}} \approx \cos(\alpha)\frac{\partial w}{\partial\bar{s}} \,.
    \label{eq:wave eq}
\end{equation}
This is just an advection equation (or first-order linear wave equation) for the curvature, so \textit{any} non-uniform initial curvature profile $w_0(\bar{s})$ will simply propagate through the filament at a constant velocity $\cos(\alpha)$ to the filament's tail \cite{Whitham2011}. The precise details of what happens to the profile when it reaches the tail are determined by the missing boundary condition but for sufficiently long filaments, this should have little effect on the behavior in the bulk.

Similarly, if only the second term on the right-hand side of Eq.~(\ref{eq:w equation limit}) contributes significantly, i.e.
\begin{equation}
    \frac{\partial w}{\partial\bar{t}} \approx 
    - g_B\frac{\partial^2}{\partial \bar{s}^2}\left[\frac{\partial^2 w}{\partial \bar{s}^2}-\frac{1}{3}w^3\right] \,,
    \label{eq:Cahn-Hilliard eq}
\end{equation}
then the curvature evolves according to a Cahn-Hilliard-like equation in one spatial dimension, with a single-well quartic potential $V(w) = w^4/12$ \cite{LiviPolitiBook}. Typical usage of the Cahn-Hilliard equation for  modeling phase separation employs a double-well quartic potential $V(w) = (w-1)^2(w+1)^2$. Here, the presence of a single-well potential in Eq.~(\ref{eq:Cahn-Hilliard eq}) allows us to think of it as a model for fluid homogenization. Physically, this means that Eq.~(\ref{eq:Cahn-Hilliard eq}) drives the filament towards a uniform curvature, the value of which is determined by the missing boundary conditions.

A final simplifying circumstance to consider is the one in which the first term on the right-hand side of Eq.~(\ref{eq:w equation limit}) is dominant. The contribution of this term is substantially more complicated than the previous two as it couples the strain and curvature such that this term cannot be studied in isolation. Accordingly, little can be said about it in full generality. Suppose however that the curvature $w(\bar{s})$ throughout the filament is small. In this case, Eq.~(\ref{eq:u equation limit}) can be approximated by
\begin{equation}
    \frac{\partial^2\delta u}{\partial\bar{s}^2}=\frac{\sin(\alpha)}{g_S}w+O(w^2) \,,
\end{equation}
thus $\delta u(\bar{s})\sim O(w)$. Upon substitution into the first term of Eq.~(\ref{eq:w equation limit}), we find that the first term contributes at order $O(w^2)$ and thus, in this small curvature limit, can be neglected relative to both the second and third terms which contain linear contributions of order $O(w)$. Accordingly, only in regions of large curvature does the strain substantially modify the curvature evolution. 

To develop intuition for how the first term behaves in the presence of large curvature, consider the particularly simple constant profile $w(\bar{s}) = w_0$. For this profile, Eq.~(\ref{eq:u equation limit}), with boundary conditions given by Eq.~(\ref{eq:BC 1 limit}), can be solved exactly to give the strain
\begin{multline}
    \delta u = -\frac{\sin(\alpha)}{g_{S}w_{0}}\Bigg[1 + \\
     \left(\frac{g_{B}w_{0}^{3}}{\sin(\alpha)}-1\right)\frac{\sinh(w_{0}\bar{s})+\sinh(w_{0}(1-\bar{s}))}{\sinh(w_{0})}\Bigg] \,,
\end{multline}
thus the first term of Eq.~(\ref{eq:w equation limit}) becomes
\begin{multline}
    g_{S}[2w\delta u'+w'\delta u]' = \\
    -2\sin(\alpha)w_0^2\left(\frac{g_Bw_0^3}{\sin(\alpha)}-1\right)\times \\ \frac{\sinh(w_{0}\bar{s})+\sinh(w_{0}(1-\bar{s}))}{\sinh(w_{0})} \,.
\end{multline}
The only factor in this expression which can change sign is $(g_Bw_0^3/\sin(\alpha)-1)$. Accordingly, if $w_0$ is large such that $g_Bw_0^3/\sin(\alpha) > 1$, this term will drive a reduction in the curvature. In contrast, if $w_0$ is small such that $g_Bw_0^3/\sin(\alpha) < 1$, this term will drive a growth in curvature. The $\bar{s}$ dependence in this expression causes the drive towards $g_Bw_0^3/\sin(\alpha) = 1$ to be non-uniform with the filament ends experiencing greater driving than the interior. Comparing with Eq.~(\ref{eq:w0 large g_S}), we thus find that it is this first term which drives the filament towards the non-trivial constant curvature solutions!

Consolidating the above analysis, we find that for curves in which the strain $\delta u$ is small, the curvature profile evolves by balancing three processes -- a wave-like propagation of the curvature profile towards the filament tail, a diffusive drive towards uniform curvature and a strain-driven tendency towards the particular curvature which constitutes the constant curvature of the non-trivial stationary state. It is the complicated interplay between these processes which determines whether a given initial condition will evolve towards a straight, unstretched configuration, a uniformly compressed, constant curvature configuration, or some other state.

\section{\label{sec:Simulations}Simulations}

To determine the region of validity of our theory, we use \textit{Mathematica} \cite{SI-mathematica-notebook} to simulate filaments subjected to Eq.~(\ref{eq:overdamped evolution}) with forces given by Eqs.~(\ref{eq:FS}) through (\ref{eq:FS at L}) and Eqs.~(\ref{eq:FB at 0}) through (\ref{eq:FA}), all nondimensionalized by Eqs.~(\ref{eq:nondim s}) through (\ref{eq:nondim k}).

Since the equations are deterministic, each simulation is entirely determined by the three dimensionless parameters $g_S$, $g_B$ and $\alpha$ along with the initial condition $\bar{r}_0(\bar{s})$. This space, however, is enormous, particularly since both the curvature $w(\bar{s})$ and the stretching $u(\bar{s})$ of the initial condition need not be uniform and can be varied arbitrarily. Accordingly, we do not make any attempt to fully characterize the space of possible behaviors of filaments. Rather, we restrict our attention to simulations which are initialized with filaments of constant curvature $w(\bar{s}) = w_0$ and stretching $u(\bar{s}) = u_0$ given by Eqs.~(\ref{eq:u0 from w0}) and (\ref{eq:polynomial}), which are predicted to be stationary in time. Below, we report on the long-time shape of such filaments and the circumstances under which they remain stationary.

Eq.~(\ref{eq:overdamped evolution}) can be iterated in time in a variety of ways, the simplest being explicit methods such as Euler's method or Runge-Kutta methods \cite{MazumderBook}. Unfortunately, for our equations, these methods require extremely small nondimensional time steps to remain numerically stable, typically of the order $\delta\bar{t} \sim 10^{-14}-10^{-12}$. Such time steps are too small to obtain results over reasonable time periods $\bar{t}_\textrm{max} \sim 10^0 -10^1$. Accordingly, instead, an implicit Euler method was implemented in which the filament configuration $\bar{r}_{\bar{t}+\delta\bar{t}}(\bar{s})$ at time $\bar{t}+\delta\bar{t}$ is determined from the filament configuration $\bar{r}_{\bar{t}}(\bar{s})$ at time $\bar{t}$ by numerically solving the equation
\begin{equation}
    \bar{r}_{\bar{t}+\delta \bar{t}}(\bar{s})=\bar{r}_{\bar{t}}(\bar{s})+\bar{F}[\bar{r}_{\bar{t}+\delta \bar{t}}(\bar{s})]\delta \bar{t} \,. \label{eq:implicit method}
\end{equation}
Though with this method, each time-step can take one or two orders of magnitude more computation time than the explicit methods, numerical stability can be achieved with time-steps typically in the range $\delta\bar{t}\sim10^{-4}-10^{-2}$, making the proposition worthwhile. To minimise the number of computations needed, an adaptive time-step algorithm was implemented. At each step, we attempt to calculate the new state of the filament using the current time-step. If a solution can be found with this time-step, it is increased (up to a maximum of \mbox{$\delta\bar{t}_\mathrm{max}=10^{-2}$)} until a solution to Eq.~(\ref{eq:implicit method}) can no longer be computed. Similarly, if we fail to compute a solution with the current time-step, it is decreased until a solution is obtained. Though this results in wasted computations since some successful computations are not used if a larger time-step is successful and failed computations are occasionally attempted, the number of time-steps needed is still smaller than just using the smallest always numerically stable time-step. 

To approximate a continuous filament, we
typically discretised the filaments into 250 points such that neighbouring points were separated by a distance $\delta\bar{s}=1/250=4\times10^{-3}$. Derivatives were computed using ordinary first-order central finite-difference schemes \cite{MazumderBook, Fornberg1988} such that they have a precision of order $O(\delta\bar{s}^2)$. Derivatives at the end points were computed using second-order one-sided central difference schemes, so that they  have the same precision. While it is possible to use a finer discretisation, this typically had no noticeable effect on the results while coming at the expense of requiring smaller time-steps and thus longer computation times. On the other hand, if the discretisation is made too small, finite size effects introduce error into the simulations. Our choice of 250 points per filament represents a reasonable compromise between these considerations.

In extreme circumstances however, such as near the boundary between stable and unstable conformations, finer discretizations can be required to determine whether a configuration is actually stable. Accordingly, for the points near these boundaries, the filaments were divided into 500 points such that $\delta \bar s=1/500=2\times10^{-3}$.

While many simulations can be successfully run with the above method, some particular parameter choices can result in configurations whose curvature at one of the end-points tends to infinity, that is, after a finite time, they develop a singularity in the curvature. In such cases, the adaptive time-step is forced to shrink to extremely small values as the singular time is approached. In the interest of saving computation time, simulations in which the time-step is shrunk to beneath $\delta\bar{t}\sim10^{-8}$ are ended early. Similarly, simulations which remain in a stable shape over a time interval $\Delta\bar{t} = 2$ are assumed to remain that way in perpetuity and are also ended early. If neither of these eventualities occur, simulations are ended after a time period of $\bar{t}_\mathrm{max} = 8$.

As predicted by our theory, straight, unstretched filaments are always stable and simply translate in space along straight trajectories. Occasionally, an initially curved state will destabilize into a roughly straight configuration, after which it will adopt the trivially straight conformation and trajectory. 

\begin{figure}
    \centering
    \includegraphics[width=1.0\linewidth]{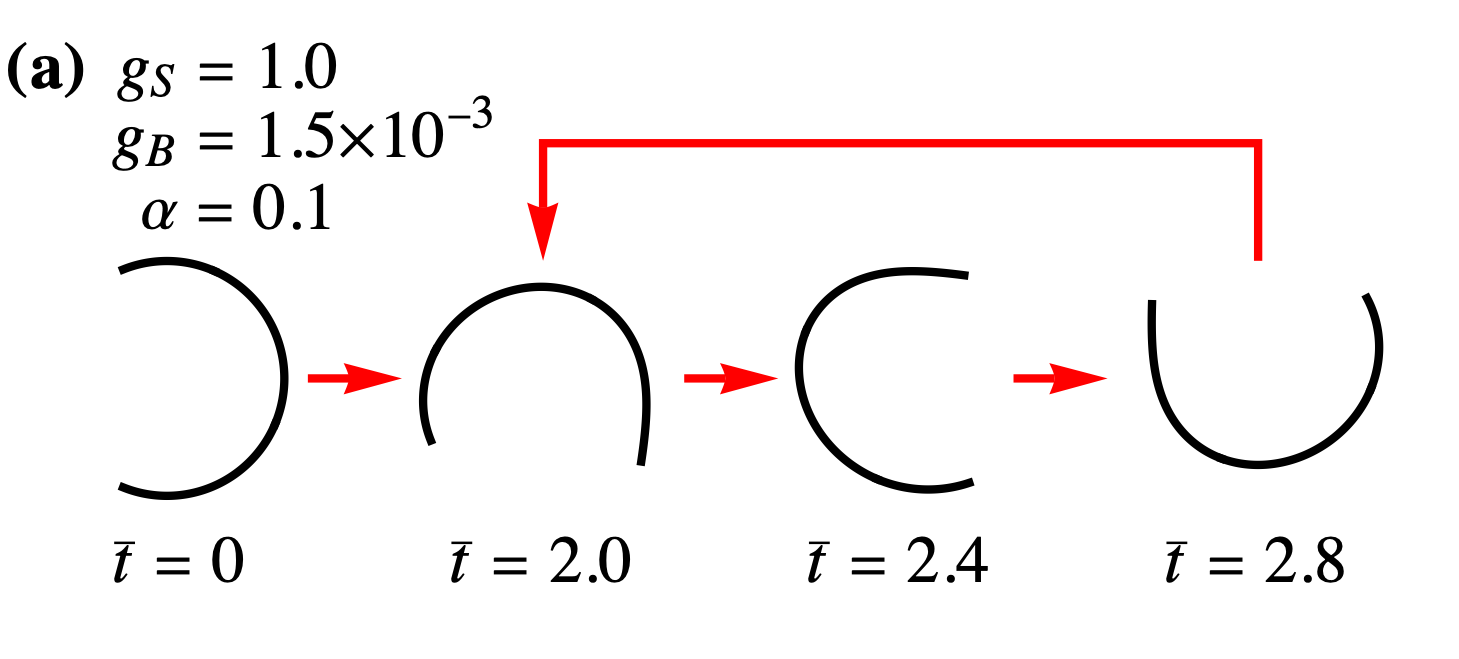}
    
    \vspace{0.08 cm}
    
    \includegraphics[width=1.0\linewidth]{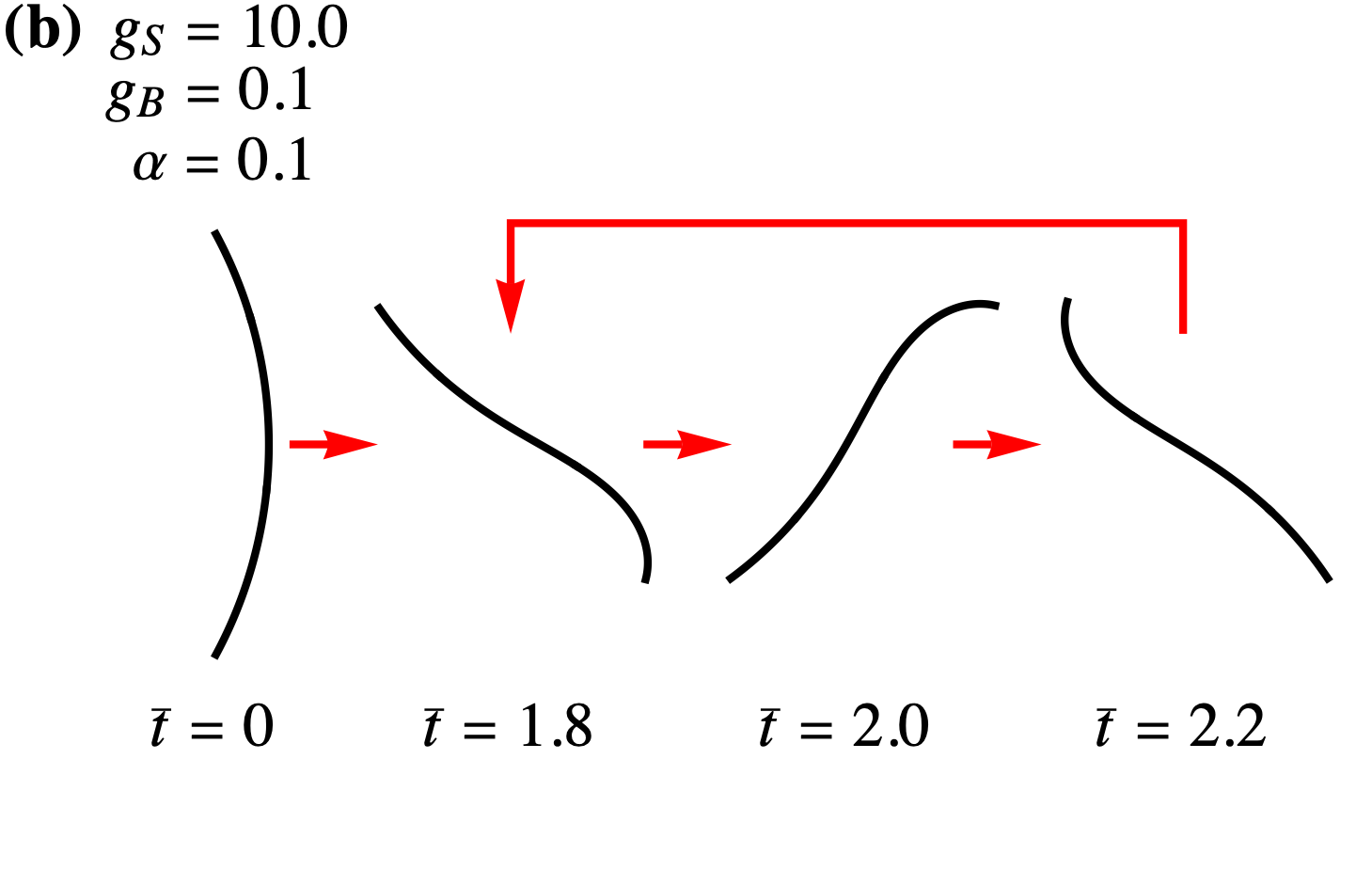}
    
    \caption{Nonuniform, stable states observed in simulations. \textbf{(a)}~A u-shaped filament, initially uniformly curved, which mostly maintains that shape by evolving into a configuration whose variation in curvature along its length is $\sim10\%$. This u-shape then rotates rigidly. \textbf{(b)}~A uniformly gently curved filament which evolves into a hook-shape which curves in both directions. This hook-shape is also stable over time,  rotating rigidly. Animations of these filaments’ evolution are shown in the Supplemental Material Movie S1 \cite{SI}.}
    \label{fig:filament shapes}
\end{figure}

More interesting are the various kinds of curved configurations that can be observed. Fig.~\ref{fig:filament shapes} shows examples of typical filament behavior observed in simulations which do not encounter singularities. The most common behavior for $\alpha=0.1$, shown in Fig.~\ref{fig:filament shapes}(a), are filaments which mostly maintain their shape. These are u-shaped (or sometimes o-shaped) initial conditions which maintain that shape qualitatively, although their initial uniform curvature is modified into a curvature which varies along the filament and deviates by up to $\sim10\%$ from its mean value. These u-shaped configurations are stable in time, merely rotating. In contrast, Fig.~\ref{fig:filament shapes}(b) shows an example of a gently curved filament which evolves into a very different shape, namely a hook-shape which curves in both directions. Such configurations are observed much more rarely, possibly because most initial conditions which would evolve into them develop a curvature singularity first. Like the u-shaped configurations, the hook-shaped configurations are stable in time, again merely rotating. Animations of these filaments' evolution are shown in  Supplementary Material Movie S1 \cite{SI}.

While we have observed such states for small $\alpha$, matters are very different for large $\alpha$. For example, for $\alpha=1$, stable u-shaped configurations are \textit{never} observed; instead, a plethora of time-dependent dynamic configurations are encountered
which never settle into a stable conformation. Though similar time-dependent states can occasionally be observed with $\alpha=0.1$, these tend to be an artifact of the numerical discretization and disappear with sufficiently fine discretizations. In contrast,  for $\alpha=1$, these dynamic states remain even upon refinement of the discretization. We thus find that for $\alpha=1$, there are no stable constant curvature solutions, at least at the computational scales we are able to access.

This is interesting because our linear stability analysis found that the size of $\alpha$ has little to no effect on the linear stability of the constant curvature configurations. 

Assuming the validity of our simulations, we are forced to conclude that increasing the size of $\alpha$ introduces strong nonlinear destabilizing effects, not accounted for by our theory.

\begin{figure}
    \centering
    \includegraphics[width=1.0\linewidth]{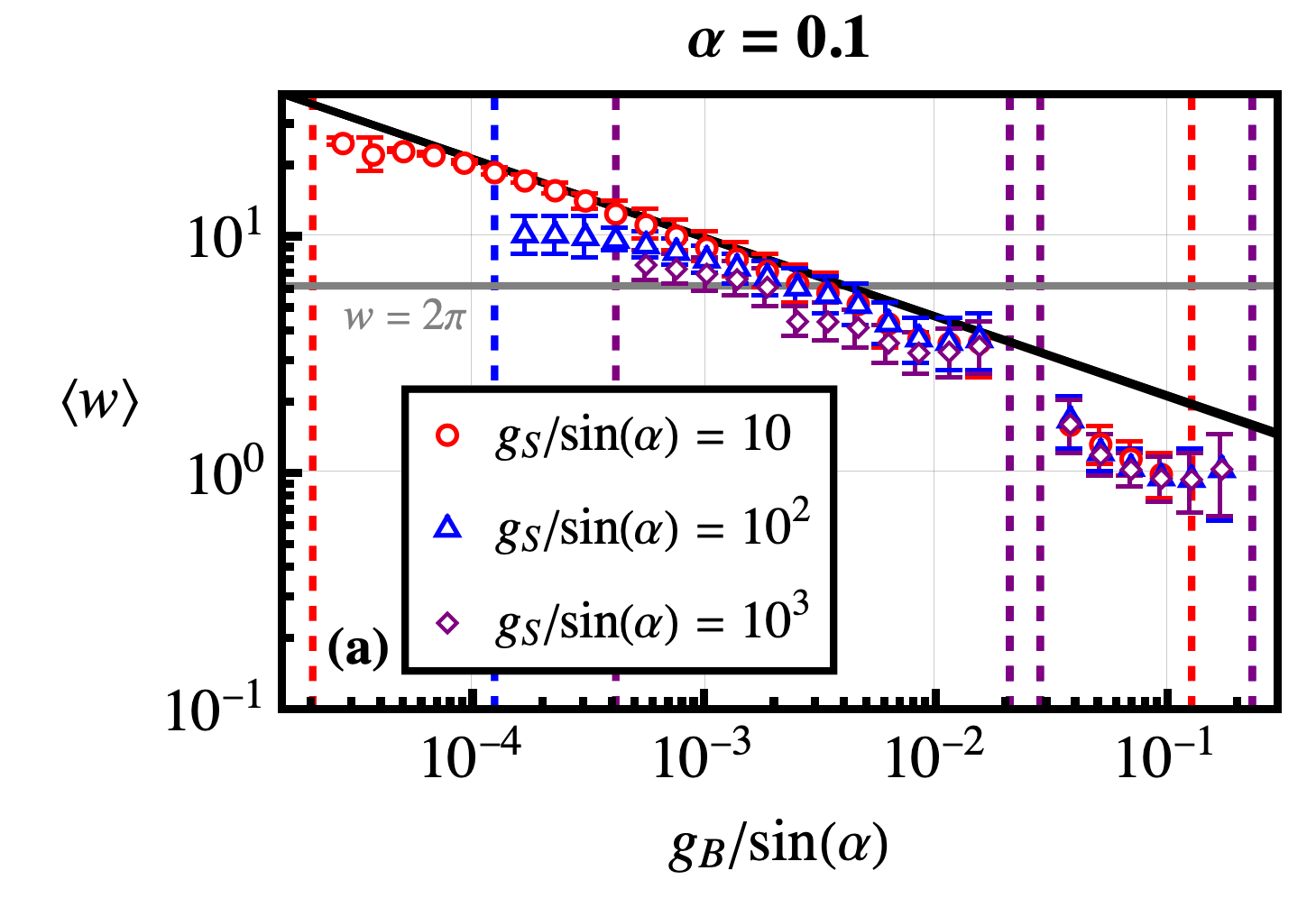}
    
    \vspace{-0.2 cm}
    
    \includegraphics[width=1.0\linewidth]{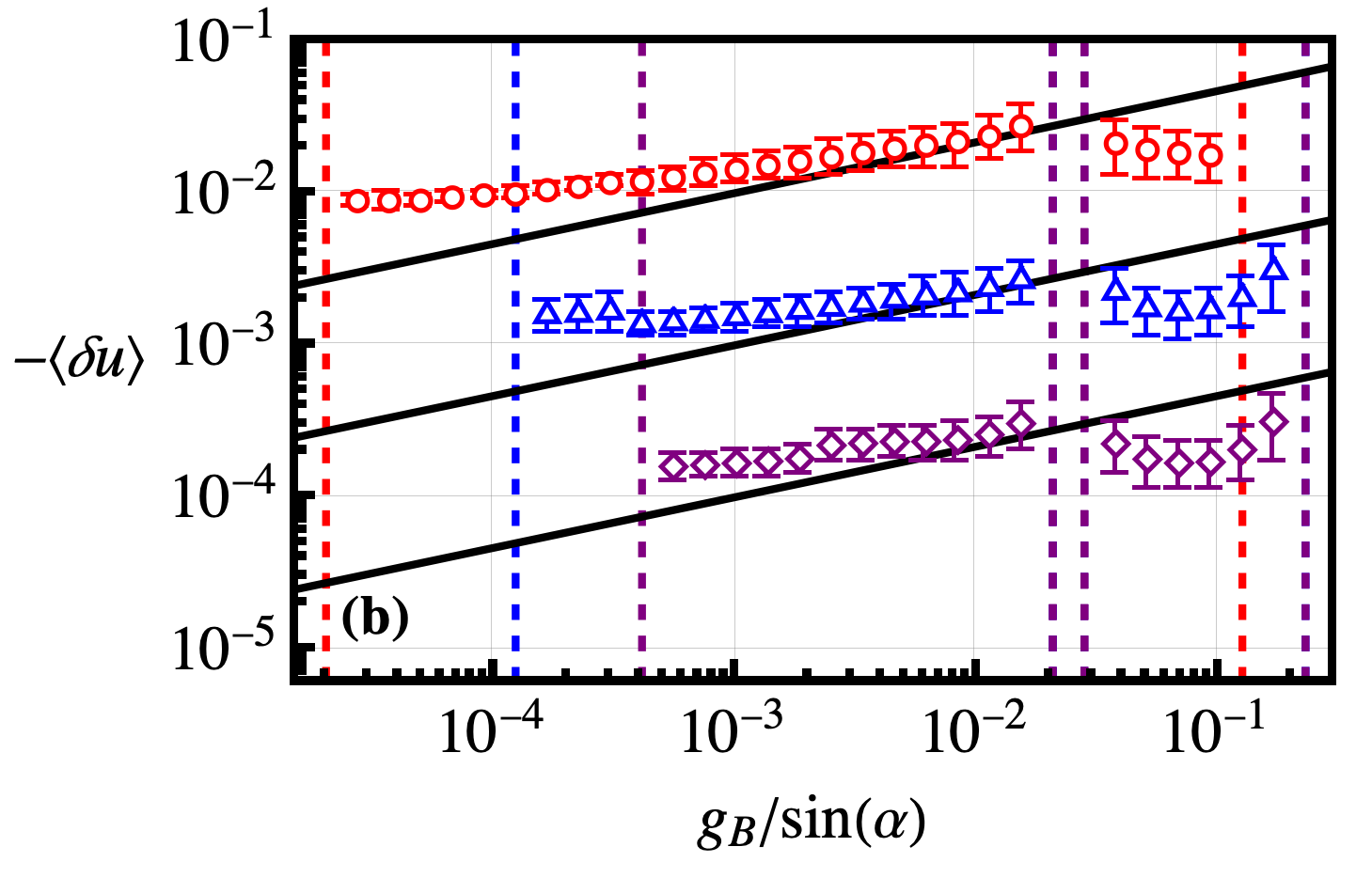}

    \vspace{-0.2 cm}

    \includegraphics[width=1.0\linewidth]{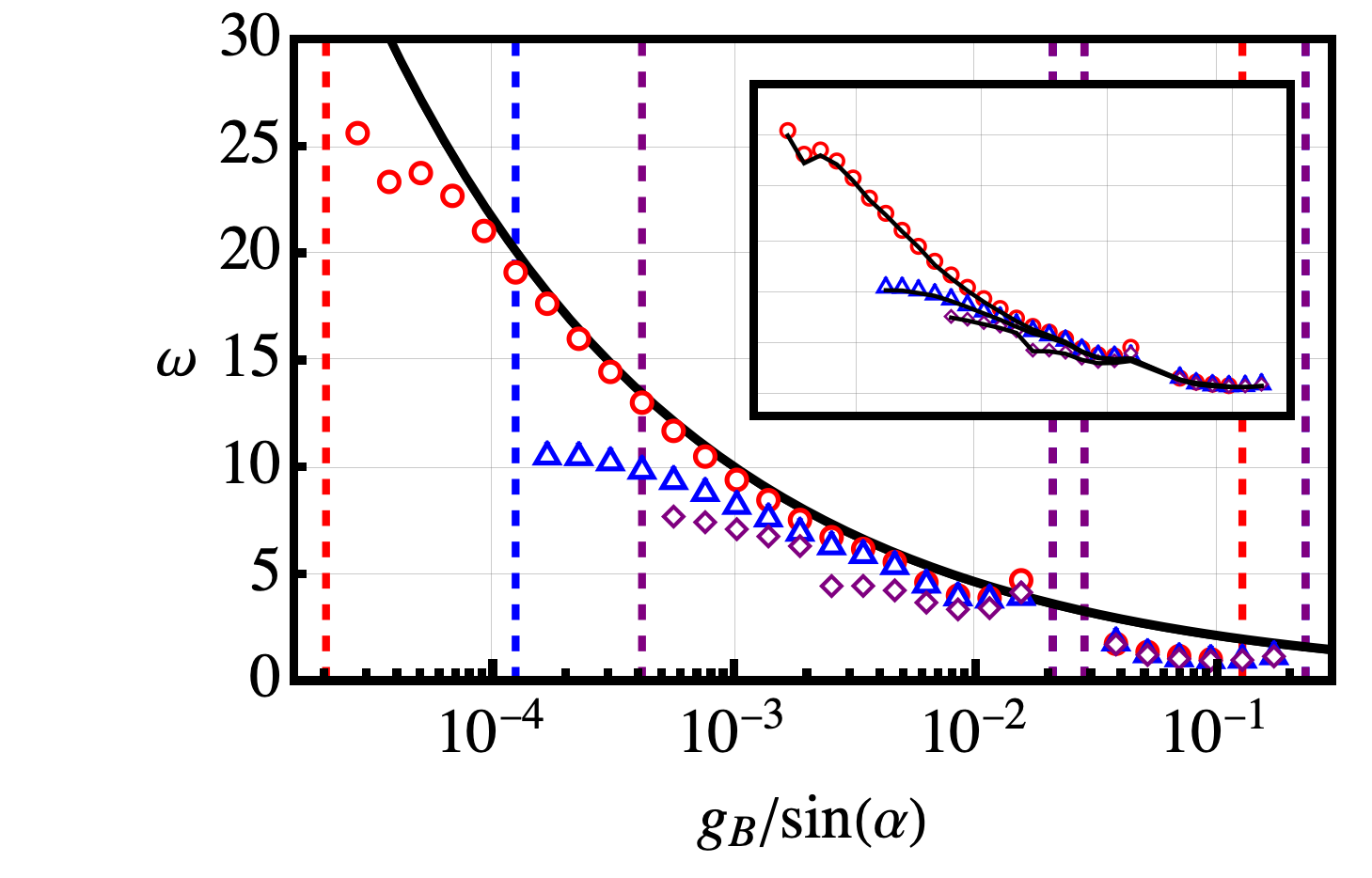}
    
    \caption{Quantitative comparison of uniformly curved states in simulation and theory. For $\alpha=0.1$, the \textbf{(a)} mean curvature $\left< w \right>$, \textbf{(b)} mean compression $\left<\delta u \right>$ and \textbf{(c)} rotation rate $\omega$, of u-shaped filaments as a function of the bending rigidity $g_B$. In each plot, data is plotted for $g_S/\sin(\alpha)$ equal to $10$ (red circles), $10^2$ (blue triangles) and $10^3$ (purple diamonds). The error bars in plots \textbf{(a)} and \textbf{(b)} correspond to one standard deviation. The solid black lines in each plot correspond to the theoretical value predicted by Eqs.~(\ref{eq:u0 from w0}), (\ref{eq:polynomial}) and (\ref{eq:rotation rate}), the first two of which are also the curvature $w$ and strain $\delta u$ of the initial condition; in  \textbf{(a)} and \textbf{(c)} the three solid black curves nearly coincide as $g_S$ affects them very little. The inset in \textbf{(c)} shows the same data points as the main figure but with solid lines calculated using Eq.~(\ref{eq:rotation rate}) with $w_0=\left<w\right>$ and $u_0=\left<u\right>$. Dashed vertical lines correspond to simulations that did not converge to a u-shape, either because they reached a singularity in curvature or because they destabilized into another state.
    }
    \label{fig:simulation plots}
\end{figure}

In Fig.~\ref{fig:simulation plots}, we present the mean curvature $\left<w\right>$, the mean strain $\left<\delta u\right>$ and the rotation rate $\omega$ of u-shaped configurations as a function of the bending rigidity $g_B$ for $\alpha=0.1$. Data is presented for three values of the stretching rigidity, such that $g_S/\sin(\alpha)\in\{10,10^2,10^3\}$. In each plot, the solid black lines correspond to the theoretical values predicted to be stable in time by Eqs.~(\ref{eq:u0 from w0}), (\ref{eq:polynomial}) and (\ref{eq:rotation rate}). In Figs.~\ref{fig:simulation plots}(a) and (b), these solid black lines also correspond to the initial curvature, $w$, and strain, $\delta u$, of each simulated filament. Though each plot has three distinct solid black lines for the three different values of $g_S$ considered, in Figs.~\ref{fig:simulation plots}(a) and (c), these lines fall almost entirely on top of each other giving the impression of a single solid black line. The fact that all data points in Figs.~\ref{fig:simulation plots}(a) and (b) fall relatively close to the solid lines suggests that filaments maintain a shape relatively close to, though not identical with, their initial configurations. Similarly, the fact that points closer to the solid lines in Figs.~\ref{fig:simulation plots}(a) and (b) also fall close to the solid line in Fig.~\ref{fig:simulation plots}(c) suggests that Eq.~(\ref{eq:rotation rate}) accurately predicts the rotation rate of a u-shaped filament. Indeed, the large deviation of some points from the solid lines in Fig.~\ref{fig:simulation plots}(c) is almost entirely due to the fact that the corresponding curvatures of these points have slightly larger deviation from the predicted value, as shown in Fig.~\ref{fig:simulation plots}(a). Were the empirical curvature, $\left<w\right>$, to be used, Eq.~(\ref{eq:rotation rate}) would correctly predict the rotation rate of all u-shaped filaments with high accuracy. This is demonstrated in the inset of Fig.~\ref{fig:simulation plots}(c) whose data points are the same as the main figure but whose solid black lines indicate the predicted rotation rate using Eq.~(\ref{eq:rotation rate}) with $w_0$ and $u_0$ being taken as the measured mean curvature $\left<w\right>$ and mean stretching factor $\left<u\right>$.

The error bars in Figs.~\ref{fig:simulation plots}(a) and (b) correspond to the standard deviation of the curvature, $w$, and the strain, $\delta u$, respectively. As can be seen, these error bars are relatively small, typically no more than $10\%$ of the value of the mean curvature, $\left<w\right>$, and mean strain, $\left<\delta u\right>$. In this sense, the configurations can be considered u-shaped, even if they are not uniformly curved.

The vertical dashed lines in Fig.~\ref{fig:simulation plots} correspond to filaments which were simulated but did not retain a u-shape, either because a singularity in the curvature occurred or because they destabilized and ultimately converged to a different shape (straight or hook-shaped). For each choice of $g_S$, a window of stability for u-shaped curves is found around the line $w=2\pi$ [the gray horizontal line in Fig.~\ref{fig:simulation plots}(a)], which corresponds to end-to-end wrapping. Note, however, that because of the small variation in the curvature along the filament, initially wrapped filaments tend to become slightly spiral and thus in practice tend to avoid issues of self-intersection. Additionally, for every choice of $g_S$ tested, the window of stability is punctured by a small island of instability around $g_B/\sin(\alpha)\sim2.5\times10^{-2}$. This qualitatively, if not quantitatively, resembles the picture predicted in Sec.~\ref{sec:Stability Analysis} by our linear stability analysis. There, we predicted that instability would occur for sufficiently small bending rigidities $g_B$ in the region of filament wrapping but that this instability would come in bands as discrete modes are alternately stabilized and destabilized with varying bending rigidity. While we do not see as many bands as predicted by the linear stability analysis, we believe that the banding feature predicted by that analysis is indeed qualitatively manifesting in these simulations. In contrast, our stability analysis also predicted stability for arbitrarily large bending rigidities (so long as constant curvature solutions continue to exist) and this is manifestly not the case in the simulations, which eventually destabilize for sufficiently large bending rigidities $g_B$. We already conjectured that such filaments would not be physically realizable, however, as the constant curvature solutions to our equations with large bending rigidities $g_B$ have large strains $|\delta u|$ and are thus highly compressed. Our simulations corroborate that conjecture through the absence of stable states with curvatures matching the small-curvature solution to Eq.~(\ref{eq:polynomial}), which require unrealistic compression. More precisely, Fig.~\ref{fig:simulation plots}(b) shows that for each data set, the magnitude of the strain $|\delta u|$ grows with increasing $g_B$ until it eventually stops growing. This can be construed as the system reaching the maximum strain the filaments can physically support and thus any constant curvature solution which requires a strain greater than this can not be realized.

\section{\label{sec:Discussion}Discussion}

In this paper, we have shown how to directly study the shape of an evolving filament in a frame that translates and rotates with it. This has particular value for active filaments which can deform under their own active forces. 
%The central thesis of this paper has been that in a homogeneous isotropic medium, the location of a spatially extended body is far less important than its shape and thus it is illuminating to study the evolution of the body's shape as directly as possible.
Motivated by the chiral character of microtubules and their apparent shape multi-stability in gliding assay experiments, we have developed a theory to predict the shapes and behaviors of chiral active elastic filaments, propelled at each point at an angle $\alpha$ relative to the filament's tangent vector at that point. 

Our theory has generated a number of predictions which we have computationally validated to varying degrees, the most important and compelling being that chiral active elastic filaments can indeed exhibit shape multi-stability. Specifically, our theory predicts that chiral active forcing along with classical elasticity are sufficient to endow active elastic filaments with both straight, unstretched states as well as uniformly curved, compressed conformations in a portion of parameter space.
\added{
As shown by Eq.~(\ref{eq:F total curved shape}), these curved conformations of filaments with no intrinsic curvature of their own can be realized when the normal component of the total force vanishes at every point along the filament. The overdamped dynamics that we assume are a necessary condition for this stability but not a sufficient one, as demonstrated by the fact that some curved conformations are in fact unstable.
}

Our simulations have shown that for small propulsion angles $\alpha$, 
\removed{this is indeed the case}
\added{shape multi-stability occurs} %
in exactly the region of parameter space predicted by our theory. In addition to these conformations, our simulations have shown that other states exist that are nonuniformly curved. While our theoretical analysis does not predict these more complicated states, it also does not rule them out. High-order nonlinear partial differential equations such as Eqs.~(\ref{eq:u equation}) and (\ref{eq:w equation}) are notoriously difficult to study, and our analysis has highlighted that the difficulty is exacerbated by having an ``incomplete'' set of boundary conditions [Eqs.~(\ref{eq:BC1 non dim}) and (\ref{eq:BC2 non dim})] to work with. 

Because of the size of the parameter space, we have made no effort to fully determine and categorize the space of possible conformations. We have, however, shown that chiral active forces introduce a rich variety of behaviors, suggesting that a more thorough exploration of parameter space  will likely yield yet more intriguing motions.

Beyond predicting the existence of non-trivial filament shapes, we have made a first attempt at predicting their stability through a linear stability analysis. This, unfortunately, predicted that highly compressed uniformly-curved states would be stable, even though such states are clearly not physically realizable. Our simulations confirm that such states cannot be realized. This discrepancy likely points to limitations on the range of validity of the force-free boundary condition assumption, given in Eq.~(\ref{eq:Force free BC}),  under which these states were found analytically. 
% Accordingly, the linear stability analysis fails to predict the instability of these states because of a ``garbage in, garbage out'' (GIGO) principle. 

In addition, the linear stability analysis predicted that stability would be nearly unaffected by the size of the angle $\alpha$. Our simulations have shown this to be far from the case, with large values of $\alpha$ completely destabilizing all uniformly curved, compressed states. This discrepancy is likely due to the limits of the linear stability approach, as our equations are highly nonlinear and these nonlinearities clearly rise in importance as $\alpha$ is increased. Capturing this feature of our equations would therefore require a full nonlinear stability analysis, which is beyond the scope of this work.

Despite these failures, the linear stability analysis does make two predictions which are qualitatively if not quantitatively confirmed by our simulations. First, it predicts that sufficiently flexible filaments will not be able to maintain a uniformly-curved shape, i.e.\ sufficiently small bending rigidities will ultimately destabilize such solutions. Second, it predicts that this destabilization will occur in ``discrete steps'' with islands of instability within the window of stability. Both of these phenomena are indeed observed in our simulations, showing that  the linear stability analysis is informative, although limited. Indeed, since our simulations found no stable  filament conformations which were predicted to be unstable, linear stability can be construed as a necessary though not sufficient condition for stability.

Finally, the approach we have used in this paper is readily generalizable to more complicated systems. For instance, thermal or other stochastic noise sources can be readily incorporated, as can external force fields. 
\added{
Direct comparison can then be made with Brownian dynamics simulations of semi-flexible bead-spring chains with off-tangent self-propulsion \cite{athani2025gliding}.
}%
Accordingly, we believe that the theoretical framework developed here for actively propelled filaments is a promising tool for predicting and explaining the behavior a broad range of systems beyond microtubule gliding assays, including other cytoskeletal filaments, synthetic polymers, and potentially  the crawling of worms on surfaces \cite{bau2015worms}. Our findings also emphasize the richness of filament behaviors arising from off-tangent propulsion, raising the possibility that systems of many such filaments may offer a route to  mechanically switchable active matter.

\appendix

\section{Variation of the Stretching and Bending Hamiltonians\label{sec:Appendix}}

In Sec.~\ref{sec:Elastic Filaments}, we identify the Hamiltonians for the stretching and bending forces of a one-dimensional curve~$\vec{r}(s)$
\begin{align}
    H_S[\vec{r}(s)] & = \frac{1}{2} \kappa_{S} 
    \int_0^L ds\, (|\vec{r}\,'(s)| - 1)^2 \,, \\
    H_B[\vec{r}(s)] & = \frac{1}{2} \kappa_B 
    \int_0^L ds\, \mathsf{k}(s)^2 \,.
\end{align}
Upon varying the curve by a small amount 
\begin{equation}
    \vec{r}(s)\rightarrow\vec{r}(s)+\delta\vec{r}(s) \,,
\end{equation}
these Hamiltonians will also vary
\begin{equation}
    H[\vec{r}(s)]\rightarrow H[\vec{r}(s)] + \delta H
\end{equation}
where we have denoted $\delta H = H[\vec{r}(s)+\delta\vec{r}(s)] - H[\vec{r}(s)]$. As it can be tricky to determine the $\delta H$ directly, it is helpful to build up a library of variations of the various physical quantities to assist with this.

We begin with the stretching factor $|\vec{r}\,'(s)|$,  for which we can write
\begin{align}
    |\vec{r}\,'(s)| 
    & \rightarrow |\vec{r}\,'(s) 
    + \delta\vec{r}\,'(s)| \nonumber\\
    & = |\vec{r}\,'(s)|\left[1+\frac{\vec{r}\,'(s)\cdot\delta\vec{r}\,'(s)}{|\vec{r}\,'(s)|^{2}}\right]  \nonumber\\
    & = |\vec{r}\,'(s)|+\hat{T}(s)\cdot\delta\vec{r}\,'(s) \,.
\end{align}
Here and in the following, we have dropped terms of order $|\delta\vec{r}(s)|^2$ and higher. 

With this in hand, it is now easy to compute the variation of the unit tangent vector $\hat{T}(s)=\vec{r}\,'(s)/|\vec{r}\,'(s)|$ as
\begin{align}
    \hat{T}(s)
    &\rightarrow\frac{\vec{r}\,'(s)+\delta\vec{r}\,'(s)}{|\vec{r}\,'(s)+\delta\vec{r}\,'(s)|}\nonumber\\
    &=\hat{T}(s)+\frac{\delta\vec{r}\,'(s)-\left[\hat{T}(s)\cdot\delta\vec{r}\,'(s)\right]\hat{T}(s)}{|\vec{r}\,'(s)|}\nonumber\\
    &=\hat{T}(s)+\frac{\hat{N}(s)\cdot\delta\vec{r}\,'(s)}{|\vec{r}\,'(s)|}\hat{N}(s) \,,
\end{align}
where we have obtained the last line by observing that the numerator in the second line is just $\delta\vec{r}\,'(s)$ minus the tangential component of $\delta\vec{r}\,'(s)$ which is equivalent to the normal component of $\delta\vec{r}\,'(s)$.

Since the unit normal vector $\hat{N}(s)=R_{\pi/2}\hat{T}(s)$ is obtained directly from the unit tangent vector by rotation through $90^\circ$, its variation now immediately follows as
\begin{equation}
    \hat{N}(s)\rightarrow\hat{N}(s)-\frac{\hat{N}(s)\cdot\delta\vec{r}\,'(s)}{|\vec{r}\,'(s)|}\hat{T}(s) \,.
\end{equation}

Accordingly, the curvature $\mathsf{k}(s) = \hat{T}'(s)\cdot \hat{N}(s)$, which is given by the projection of the derivative of $\hat{T}(s)$ on $\hat{N}(s)$, will have variation
\begin{widetext}
\begin{equation}
    \mathsf{k}(s)\rightarrow\left[\hat{T}(s)+\frac{\hat{N}(s)\cdot\delta\vec{r}\,'(s)}{|\vec{r}\,'(s)|}\hat{N}(s)\right]'\cdot \left[\hat{N}(s)-\frac{\hat{N}(s)\cdot\delta\vec{r}\,'(s)}{|\vec{r}\,'(s)|}\hat{T}(s)\right]
    =\mathsf{k}(s)+\left[\frac{\hat{N}(s)\cdot\delta\vec{r}\,'(s)}{|\vec{r}\,'(s)|}\right]' \,,
\end{equation}
\end{widetext}
where we have made use of the fact that $\hat{T}\cdot\hat{N}=\hat{T}\cdot\hat{T}'=\hat{N}\cdot\hat{N}'=0$ and again neglected terms of order $|\delta\vec{r}(s)|^2$ to simplify this expression.

These results are sufficient to calculate the variation of the Hamiltonians. For the stretching Hamiltonian, it is now straightforward to obtain
\begin{equation}
    \delta H_{S}=\kappa_{S}\int_{0}^{L}ds\,(|\vec{r}\,'(s)|-1)\hat{T}(s)\cdot\delta\vec{r}\,'(s)
\end{equation}
whereas for the bending Hamiltonian, we find
\begin{equation}
    \delta H_{B}=\kappa_{B}\int_{0}^{L}ds\,\mathsf{k}(s)\left[\frac{\hat{N}(s)\cdot\delta\vec{r}\,'(s)}{|\vec{r}\,'(s)|}\right]' \,.
\end{equation}
Integration by parts now gives Eqs.~(\ref{eq:delta H_S}) and (\ref{eq:delta H_B}) of Sec.~\ref{sec:Elastic Filaments}.

% \clearpage
\bibliography{bib_file}

\end{document}